\documentclass[prb,twocolumn,10pt,aps,longbibliography,nofootinbib]{revtex4-1}

 \usepackage{verbatim}
 \usepackage{amsmath}
 \usepackage{amssymb}
 \usepackage{amsthm}
 \usepackage{dsfont}
 \usepackage{latexsym}
 \usepackage{amsfonts}
 \usepackage{epsfig}
 \usepackage{epstopdf}
 \usepackage{color}
 \definecolor{darkblue}{rgb}{0,0,.5}
 \usepackage[linktocpage, colorlinks=true, linkcolor=darkblue, citecolor=darkblue]{hyperref}
 \usepackage[all]{hypcap}

\newcommand{\C}[1]{{\cal{#1}}}

\newcommand{\abs}[1]{{\left| #1 \right|}}

\newcommand{\rl}[0]{{\rangle\langle}}

\begin{document}

\title{Fermionic reaction coordinates and their application to an autonomous Maxwell demon in the strong coupling regime}

\author{Philipp Strasberg$^1$}
\email{philipp.strasberg@uni.lu}
\author{Gernot Schaller$^2$}
\author{Thomas L. Schmidt$^1$}
\author{Massimiliano Esposito$^1$}
\affiliation{${}^1$Physics and Materials Science Research unit, University of Luxembourg, L-1511 Luxembourg, Luxembourg}
\affiliation{${}^2$Institut f\"ur Theoretische Physik, Technische Universit\"at Berlin, Hardenbergstra\ss e 36, D-10623 Berlin, Germany}

\date{\today}

\begin{abstract}
 We establish a theoretical method which goes beyond the weak coupling and Markovian approximations while remaining 
 intuitive, using a quantum master equation in a larger Hilbert space. The method is applicable to all impurity 
 Hamiltonians tunnel-coupled to one (or multiple) baths of free fermions. The accuracy of the method is in principle not 
 limited by the system-bath coupling strength, but rather by the shape of the spectral density and it is especially suited 
 to study situations far away from the wide-band limit. In analogy to the bosonic case, we call it the fermionic reaction 
 coordinate mapping.  
 As an application we consider a thermoelectric device made of two Coulomb-coupled quantum dots. We pay particular 
 attention to the regime where this device operates as an autonomous Maxwell demon shoveling electrons against the voltage 
 bias thanks to information. Contrary to previous studies we do not rely on a Markovian weak coupling description. 
 Our numerical findings reveal that in the regime of strong coupling and non-Markovianity, the Maxwell demon is often 
 doomed to disappear except in a narrow parameter regime of small power output. 
\end{abstract}

\maketitle

\newtheorem{lemma}{Lemma}[section]
\newtheorem{thm}{Theorem}[section]

\section{Introduction}

Many problems in quantum transport are modeled by an impurity Hamiltonian $H_\text{imp}$ linearly coupled to a 
bath of free fermions described via 
\begin{equation}\label{eq H imp}
 H = H_\text{imp} + \sum_k(t_k d c_k^\dagger + h.c.) + \sum_k \epsilon_k c_k^\dagger c_k.
\end{equation}
Here, $c_k^{(\dagger)}$ annihilates (creates) a fermion of energy $\epsilon_k$ in the bath, which is tunnel-coupled 
with complex amplitude $t_k$ to the system via a fermionic annihilation (creation) operator $d^{(\dagger)}$ of the 
system. The actual physical system under study is described by $H_\text{imp}$, which could be one (or multiple) quantum 
dots, molecules in mechanically controllable break junctions or nano-electro-mechanical systems, among other possible 
impurity systems. 

To treat such systems theoretically, various approximate or formally exact techniques have been developed, such as 
quantum master equations (MEs)~\cite{BreuerPetruccioneBook2002, SchallerBook2014, DeVegaAlonsoRMP2017}, the formalism of 
nonequilibrium Green's functions~\cite{StefanucciVanLeeuwenBook2013} or renormalization group 
techniques~\cite{BullaCostiPruschkeRMP2008}. Whereas MEs easily allow to treat interactions in the impurity even under 
nonequilibrium situations, their use is limited to the weak coupling, Markovian and high temperature (``sequential 
tunneling'') regime. Green's functions can overcome the latter problem, but have difficulties to treat interacting 
impurities (for instance, due to Coulomb forces). Whereas this problem can be tackled by using numerical renormalization 
group approaches, they in turn are hard to apply far away from equilibrium. 

In the first half of this article (Sec.~\ref{sec fermionic RCs}), we will introduce a technique which 
lies ``in between'' these approaches. It allows to some extent to overcome the limitations 
of the standard perturbative approach commonly applied to obtain a ME while still retaining a description in terms of 
a ME such that interactions and nonequilibrium situations can be conveniently treated. The price to 
pay is an enlarged Hilbert space, in which a suitable redefined impurity Hamiltonian $\tilde H_\text{imp}$ includes 
particularly choosen dominant degrees of freedom from the bath, which we will call {\it fermionic reaction coordinates} 
(RCs). In the context of linear bosonic reservoirs (Caldeira-Leggett or Brownian motion models), this technique has a 
longer tradition~\cite{GargOnuchicAmbegaokarJCP1985}. It has found various 
applications in the theory of open quantum systems~\cite{JeanFriesnerFlemingJCP1992, MayKuhnSchreiberJPC1993, 
PollardFriesnerJCP1994, WolfsederDomckeCPL1996, WolfsederEtAlCP1998, HartmannGoychukHanggiJCP2000, ThossWangMillerJCP2001, 
CederbaumGindenspergerBurghardtPRL2005, HughesChristBurghardtJCP2009a, HughesChristBurghardtJCP2009b, 
MartinazzoEtAlJCP2011, MartinazzoHughesBurghardtPRE2011, IlesSmithLambertNazirPRA2014, HuhEtAlNJP2014, 
BonfantiEtAlAnnPhys2015, IlesSmithEtAlJCP2016} and it is also closely related to the 
``time evolving density matrix using orthogonal polynomials algorithm'' (TEDOPA)~\cite{PriorEtAlPRL2010, 
ChinEtAlJMP2010, WoodsEtAlJMP2014, WoodsCramerPlenioPRL2015, WoodsPlenioJMP2016, RosenbachEtAlNJP2016}. We remark that, 
although it shares many similarities with the bosonic case, the RC mapping was not studied for fermionic reservoirs before. 
In addition, our article adds additional insights to the recent attempts to find a meaningful thermodynamic description 
beyond the weak coupling and Markovian regime~\cite{EspositoLindenbergVandenBroeckNJP2010, 
EspositoOchoaGalperinPRL2015, EspositoOchoaGalperinPRB2015, BruchEtAlPRB2016, SeifertPRL2016, StrasbergEtAlNJP2016, 
TalknerHaenggiPRE2016, KatoTanimuraJCP2016, CerrilloBuserBrandesPRB2016, WhitneyArXiv2016, NewmanMintertNazirPRE2017, 
JarzynskiPRX2017, MillerAndersPRE2017, StrasbergEspositoPRE2017, PerarnauLlobetEtAlArXiv2017, MuEtAlNJP2017, 
HaughianEspositoSchmidtPRB2018, RestrepoEtAlArXiv2017}. In particular, our work shows that techniques based on a 
redefined system-bath partition using RC mappings~\cite{StrasbergEtAlNJP2016, NewmanMintertNazirPRE2017, 
StrasbergEspositoPRE2017, RestrepoEtAlArXiv2017} turn out to be useful for fermionic reservoirs, too. 

In the second half of the article (Secs.~\ref{sec electronic MD review} and~\ref{sec electronic MD beyond weak coupling}) 
we make use of our new method to study two (spinless) Coulomb-coupled quantum dots in contact with three heat reservoirs. 
This setup is raising increasing attention within the context of quantum thermodynamics, as it provides a prototypical 
example of a thermoelectric device transporting electrons against a potential bias due to an energetic flow from a hot 
to a cold bath~\cite{SanchezJordanNanotechnology2015, BenentiEtAlPhysRep2017}. It is well-studied in the weak-coupling 
and Markovian regime, theoretically~\cite{SanchezBuettikerPRB2011, StrasbergEtAlPRL2013} as well as 
experimentally~\cite{HartmannEtAlPRL2015, ThierschmannEtAlNatNanotech2015}. 
Moreover, Ref.~\cite{StrasbergEtAlPRL2013} identified necessary conditions which guarantee that this device can be 
interpreted as an autonomous Maxwell demon (MD), i.e., a device which is capable of extracting work (in this case 
by charging a battery) due to a clever way of processing information. From this perspective the device was further 
studied theoretically~\cite{HorowitzEspositoPRX2014, KutvonenEtAlSciRep2015} and experimentally~\cite{KoskiEtAlPRL2015}. 

Unfortunately, the study in Ref.~\cite{StrasbergEtAlPRL2013} has revealed that a proper operation of the device as a 
MD requires a {\it strong} coupling to a {\it cold} reservoir and {\it structured} (i.e., non-Markovian) spectral 
densities for the hot reservoirs. These three requirements challenge the usual range 
of validity of a ME description, but the question whether the transparent interpretation of a MD also holds under these 
conditions has not been answered yet. 
By using the method of fermionic RCs, we will indeed show that the device can still be interpreted as a MD, albeit in a 
very narrow parameter regime restricted to low power output. Our results are also in qualitative agreement 
with a recent study~\cite{WalldorfJauhoKaasbjergPRB2017} where cotunneling effects were taken into account. 
Another complementary paper studies the impact of strong coupling effects for non-autonomous, i.e., 
measurement-based, feedback loops on the thermodynamic performance of a MD~\cite{SchallerEtAlArXiv2017}.

\section{Fermionic reaction coordinates}
\label{sec fermionic RCs}

In this section we develop the theory of fermionic reaction coordinates (RCs), which is useful to explore the physics of 
open systems beyond the weak coupling, Markovian and high temperature assumption. Technically speaking, this mapping 
is a unitary transformation applied to the bath Hamiltonian, which allows to separate out a particular degree 
of freedom in the bath, called the RC. The details of this transformation are reported in 
Sec.~\ref{sec technical details}. 

The hope is then that the original impurity {\it together} with the RC is only weakly coupled to a Markovian residual 
bath such that it is possible to apply standard master equation (ME) procedures to this redefined system-bath 
partition. Whether this is possible is case-dependent and will be discussed in Sec.~\ref{sec final remarks}. 

To make the paper self-contained, we also give a short review in Appendix~\ref{app ME} of the ME approach including the 
definition of energy and particle currents, which we will use in the second part of this paper. Furthermore, 
in Appendix~\ref{app benchmark} we benchmark the fermionic RC method against the exact solution of the Fano-Anderson 
model (also known as single electron transistor).

\subsection{The mapping}
\label{sec technical details}

The mapping will work whenever it is allowed to describe the interaction between an arbitrary impurity Hamiltonian 
$H_\text{imp}$ and the fermionic reservoir as in Eq.~(\ref{eq H imp}). An important quantity in the study of such 
open systems is the spectral density (SD) (also called hybridization function) of the bath, which is defined as 
\begin{equation}
 J(\omega) \equiv 2\pi\sum_k |t_{k}|^2\delta(\omega-\epsilon_{k}).
\end{equation}
It contains the complete information about the way in which the bath is coupled to the system and will be of central 
importance in the following. 

For mathematical rigour one often demands that $J(\omega)$ is strictly greater than zero for 
$\omega\in[\omega_L,\omega_R]$ and zero outside this interval where $\omega_L < \omega_R \in\mathbb{R}$ are referred to 
as cutoff frequencies~\cite{MartinazzoEtAlJCP2011}. However, as long as all quantities converge, our equations also remain 
valid if the SD decays only exponentially or polynomially. Convergence problems for infinite cutoff frequencies 
arise only when the mapping is applied iteratively (see below). Furthermore, gapped SDs (having support only at 
disconnected intervals) can be treated by applying this mapping to each sub-SD separately. 

For later reference, we start by considering the Heisenberg equation of motion for $d$ and $c_k$ (suppressing the 
time-dependence on all operators and setting $\hbar\equiv 1$ throughout): 
\begin{align}
 \dot d		&=	i[H_\text{imp},d] + i\sum_k t_k^* c_k,	\\
 \dot c_k	&=	-i\epsilon_k c_k + it_k d.
\end{align}
We Fourier transform them according to the definition 
$\hat f(z) \equiv \int_{-\infty}^\infty dt e^{izt} f(t)$ [with $\Im(z) > 0$]. This yields 
(again dropping the explicit $z$-dependence) to 
\begin{align}
 -iz\hat d	&=	i\widehat{[H_\text{imp},d]} + i\sum_k t_k^* \hat c_k,	\label{eq Heisenberg d Fourier}	\\
 -iz\hat c_k	&=	-i\epsilon_k \hat c_k + it_k \hat d.	\label{eq Heisenberg c_k Fourier}
\end{align}
After some algebra we obtain a formally exact expression for $\hat d$, which reads 
\begin{equation}\label{eq EOM d}
 -iz\hat d = i\widehat{[H_\text{imp},d]} + \frac{i}{2} W_0(z)\hat d.
\end{equation}
Here, we introduced the Cauchy transform 
\begin{equation}
 W_0(z) \equiv \frac{1}{\pi}\int_{\omega_L}^{\omega_R} d\omega \frac{J(\omega)}{\omega - z}.
\end{equation}
By the Sokhotski-Plemelj theorem this fulfills for $\omega\in\mathbb{R}$ 
\begin{equation}\label{eq useful identitz W_0}
 \begin{split}
  W_0^+(\omega)	&\equiv	\lim_{\epsilon\searrow0} W_0(\omega+i\epsilon)	\\
		&=	iJ(\omega) + \C P\int_{\omega_L}^{\omega_R} \frac{d\omega'}{\pi}\frac{J(\omega')}{\omega'-\omega},
 \end{split}
\end{equation}
where $\C P$ denotes the Cauchy principal value. 

We now perform the mapping by introducing a new set of fermionic creation and annihilation operators 
$\{C_k^{(\dagger)}\}$ via $C_k = \sum_l\Lambda_{kl} c_l$ where $\Lambda$ is a unitary matrix fulfilling 
$\Lambda\Lambda^\dagger = 1$ such that the fermionic anti-commutation relations are preserved. 
In particular, we fix the first row of $\Lambda$ by the requirement 
\begin{equation}
 \lambda_0^* C_1 = \sum_{k} t_k^*c_k,
\end{equation}
where the parameter $\lambda_0$ is fixed by the requirement $\{C_1,C_1^\dagger\} = 1$, which implies 
\begin{equation}\label{eq lambda_0}
 |\lambda_0|^2 = \sum_k |t_k|^2 = \int_{\omega_L}^{\omega_R} \frac{d\omega}{2\pi}J(\omega). 
\end{equation}
In analogy with the bosonic case we will call $C_1$ a fermionic RC or collective coordinate.
Furthermore, we demand for $l\neq 1$ and $m\neq 1$ that 
$\sum_{k} \epsilon_k \Lambda_{lk} \Lambda_{mk}^* = \delta_{lm} E_l$. 
Hence, in terms of the new fermions, the Hamiltonian~(\ref{eq H imp}) becomes 
\begin{align}
 \tilde H	=&~	H_\text{imp} + \lambda_0dC_1^\dagger + \lambda_0^*C_1d^\dagger + E_1C_1^\dagger C_1	\label{eq Hamiltonian transformed}	\\
		&+	\sum_{k}\left(T_k^*C_1^\dagger C_k + T_kC_k^\dagger C_1\right) + \sum_{k} E_k C_k^\dagger C_k\,,\nonumber
\end{align}
with $T_k = \sum_{m} \epsilon_m t_m\Lambda_{km}/\lambda_0$ and 
\begin{equation}\label{eq E_1}
 E_1 = \sum_k \frac{\epsilon_k|t_k|^2}{|\lambda_0|^2} = \int_{\omega_L}^{\omega_R} \frac{d\omega}{2\pi} \frac{\omega J(\omega)}{|\lambda_0|^2}.
\end{equation}
The complex phase of the constant $\lambda_0$ is not fixed by this procedure, but it can be fully absorbed by redefining 
the operator $C_1$, adding only an additional phase to the renormalized couplings $T_k$. We will therefore consider 
$\lambda_0$ positive in our numerical investigations. The effect of the residual modes $\{C_k\}_{k\neq1}$ is specified by 
the residual SD $J_1(\omega) = 2\pi\sum_k |T_k|^2\delta(\omega-E_k)$, 
which we now finally determine as a functional of the original SD $J(\omega)$. 

For this purpose we look again at the Heisenberg equations, but now within the transformed coordinates. After Fourier 
transformation we obtain analogously to Eqs.~(\ref{eq Heisenberg d Fourier}) and~(\ref{eq Heisenberg c_k Fourier}) 
\begin{align}
 -iz\hat d	&=	i\widehat{[H_\text{imp},d]} + i\lambda_0^*\hat C_1,	\\
 -iz\hat C_1 	&=	-iE_1\hat C_1 + i\lambda_0\hat d - i\sum_{k=2}^N T_k^*\hat C_k,	\\
 -iz\hat C_k 	&=	-iE_k\hat C_k -i T_k\hat C_1 ~~~ (k\neq 1).
\end{align}
Again, we can formally solve for $C_k$ and $C_1$ to obtain an exact expression for $\hat d$: 
\begin{equation}\label{eq EOM d RC}
 -iz\hat d = i\widehat{[H_\text{imp},d]} + i\frac{|\lambda_0|^2}{E_1 - z - \frac{1}{2}W_1(z)}\hat d.
\end{equation}
Since both Eqs.~(\ref{eq EOM d}) and~(\ref{eq EOM d RC}) are exact, they must coincide. Thus, by comparison we obtain 
\begin{equation}\label{eq nice relation Cauchy transform}
 W_1(z) = 2(E_1 - z) - \frac{4|\lambda_0|^2}{W_0(z)},
\end{equation}
and due to relation~(\ref{eq useful identitz W_0}), we get 
\begin{equation}\label{eq mapping SD}
 J_1(\omega) = \Im[W_1^+(\omega)] = \frac{4|\lambda_0|^2J(\omega)}{|W_0^+(\omega)|^2}.
\end{equation}

\begin{figure}
 \centering\includegraphics[width=0.48\textwidth,clip=true]{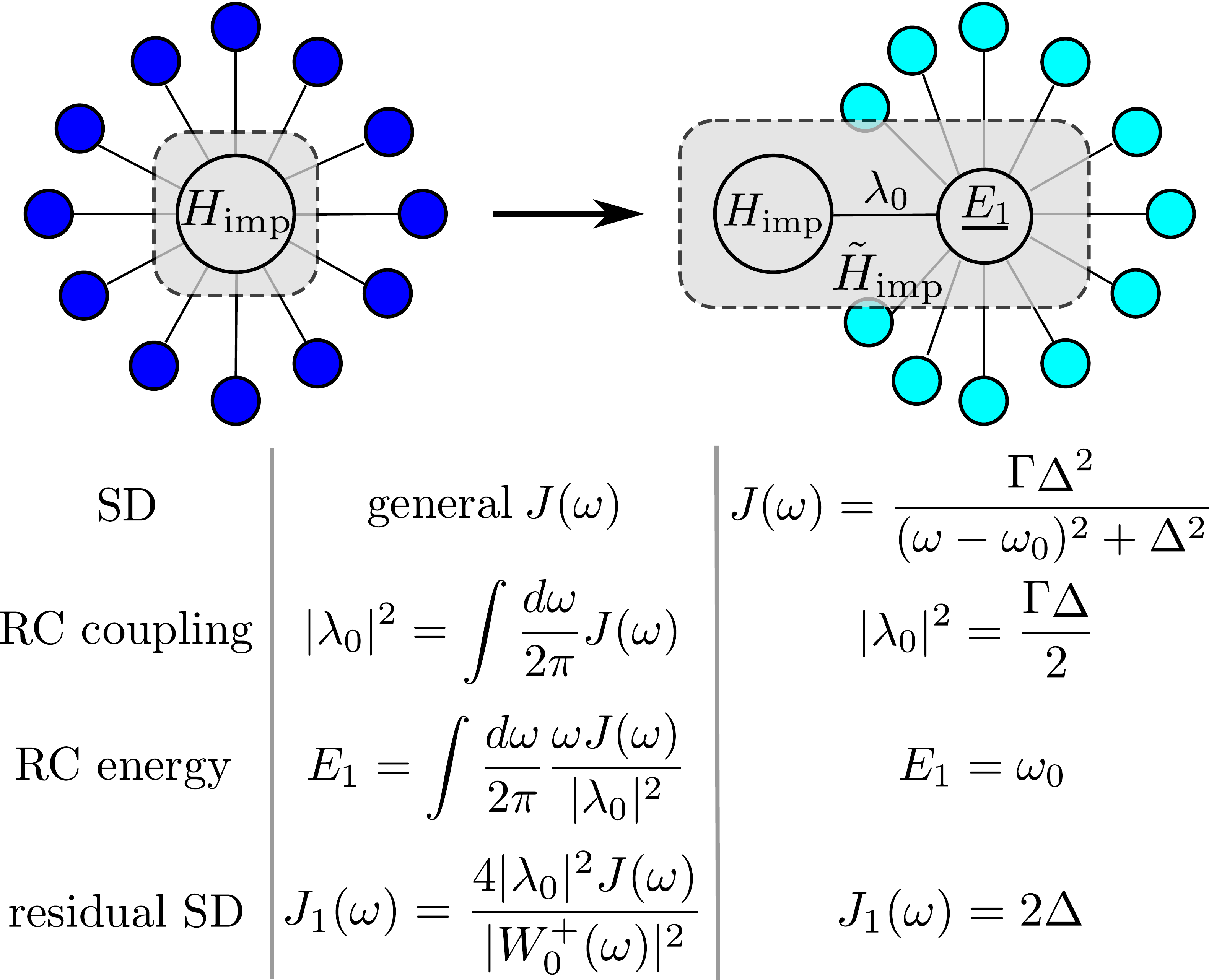}
 \label{fig RC mapping} 
 \caption{Pictorial representation of the RC mapping (top) and the equations determining the transformation for the 
 general case (middle column) and the example of a Lorentzian SD as used throughout the text (right column). 
 The shaded grey area in the sketch indicates which part is treated as the impurity whereas the remaining part is 
 treated as the bath in spirit of the ME approach outlined in Sec.~\ref{app ME}. We used different colors for the 
 free fermions in the original/residual bath to emphasize that they are not the same before and after the mapping. }
\end{figure}

This finally completes the mapping: performing a specific normal-mode transformation on the 
Hamiltonian~(\ref{eq H imp}) yields a new Hamiltonian~(\ref{eq Hamiltonian transformed}). The new 
parameters $\lambda_0$, $E_1$ and the new SD $J_1(\omega)$ are given by Eqs.~(\ref{eq lambda_0}),~(\ref{eq E_1}) 
and~(\ref{eq mapping SD}), respectively. They are all completely specified in terms of the initial SD $J(\omega)$. 
We remark that this mapping is formally exact, no approximation has been made. The mapping is summarized 
in Fig.~\ref{fig RC mapping}. 

Before proceeding, we remark that the impurity Hamiltonian $H_\text{imp}$ is allowed to be explicitly time-dependent 
without invalidating any point in the derivation above as the explicit form of $H_\text{imp}$ never entered the derivation. 
In fact, even a global time-dependence of the tunnel Hamiltonian $H_I$ of the form $\alpha(t)H_I$ does not invalidate 
any point of our derivation above as we can include the time-dependence $\alpha(t)$ in the definition of the operator 
$d'(t) \equiv \alpha(t) d$. The reason for this generality comes from the fact that the mapping is a unitary transformation 
in the bath Hilbert space alone and does not touch the system Hilbert space. 

Finally, we investigate what happens if we apply the mapping iteratively. In fact, if we define a new impurity Hamiltonian 
\begin{equation}
 \tilde H_\text{imp} \equiv H_\text{imp} + \lambda_0dC_1^\dagger + \lambda_0^*C_1d^\dagger + E_1C_1^\dagger C_1\,,
\end{equation}
a new tunnel coupling $\tilde{H}_\text{I} = \sum_k \left(T_k^* C_1^\dagger C_k + T_k C_k^\dagger C_1\right)$ and a new 
residual reservoir $\tilde{H}_\text{R} = \sum_k E_k C_k^\dagger C_k$, we see that the 
Hamiltonian~(\ref{eq Hamiltonian transformed}) has the same structure as~(\ref{eq H imp}). Thus, applying the mapping 
iteratively we obtain a chain of RCs with coupling constants $\lambda_0, \lambda_1, \dots$, energies $E_1, E_2, \dots$ 
and residual SDs $J_1(\omega), J_2(\omega), \dots$. These can be determined recursively in full analogy
\begin{align}\label{EQ:mapping_recursive}
 \abs{\lambda_{n}}^2	&=	\int_{\omega_L}^{\omega_R} \frac{d\omega}{2\pi} J_n(\omega),	\\
 E_{n+1}		&=	\int_{\omega_L}^{\omega_R} \frac{d\omega}{2\pi \abs{\lambda_n}^2} J_n(\omega),	\\
 J_{n+1}(\omega) 	&=	\frac{4 \abs{\lambda_{n}}^2 J_n(\omega)}{\left[{\cal P} \int_{\omega_L}^{\omega_R} \frac{J_n(\omega')}{\omega'-\omega} \frac{d\omega'}{\pi}\right]^2 + \left[J_n(\omega)\right]^2},
\end{align}
where we have additionally inserted Eq.~(\ref{eq useful identitz W_0}). In analogy to the bosonic 
case~\cite{MartinazzoEtAlJCP2011}, it is therefore natural to ask what is the limiting SD 
$\bar J(\omega)$ obtained after $n\rightarrow\infty$ iterations. 

Assuming that the limit exists and denoting 
\begin{equation}
 [\bar \lambda,\bar E,\bar W(z)] = \lim_{n\rightarrow\infty}[\lambda_n,E_n,W_n(z)],
\end{equation}
we obtain from Eq.~(\ref{eq nice relation Cauchy transform}) the condition 
\begin{equation}
 \bar W(z) = 2(\bar E - z) - \frac{4|\bar \lambda|^2}{\bar W(z)}
\end{equation}
with the two possible solutions $\bar W_\pm(z) = \bar E-z \pm\sqrt{(\bar E-z)^2-4|\bar \lambda|^2}$. The SD for 
$(\bar E-\omega)^2 \le 4|\bar\lambda|^2$ therefore becomes 
\begin{equation}\label{eq limiting SD}
 \bar J(\omega) = \sqrt{4|\bar \lambda|^2 - (\bar E-\omega)^2}
\end{equation}
where the requirement of positivity fixes a unique solution of the square root. Thus, the limiting SD describes a 
{\it semi-circle} with radius $2|\bar \lambda|$ centered around $\omega = E$. In fact, $\bar E$ and $|\bar \lambda|$ are 
fixed by the initial cutoff frequencies via $\bar E = (\omega_L + \omega_R)/2$ and $2|\bar\lambda| = \omega_R - \omega_L$. 
Indeed, if this were not the case, one would obtain a contradiction. This follows from Eq.~(\ref{eq mapping SD}) together 
with our initial assumption that the SD is strictly greater for $\omega\in(\omega_L,\omega_R)$ and zero outside this 
interval. We remark that our reasoning here does not allow to draw any conclusion about the behaviour of convergence. In 
particular, it could happen that the sequence of SDs does not converge (e.g., for infinite cutoff frequencies). 
The necessary conditions for convergence were studied by Woods {\it et al.}~\cite{WoodsEtAlJMP2014}. 

\subsection{Advantages of the RC mapping}
\label{sec final remarks}

The previous section described how to apply a unitary transformation to the bath such that it can be mapped to a new, 
redefined impurity coupled to a residual bath. It is different from the conventional bosonic 
mapping~\cite{MartinazzoEtAlJCP2011}. There, the final Hamiltonian is described in terms of position and momentum 
operators and has a different quadratic form, which -- rewritten with creation and annihilation operators -- displays 
counter-rotating terms which are absent in the fermionic case. See also Ref.~\cite{WoodsEtAlJMP2014} for further details 
on this point. 

We emphasize that this mapping is formally exact, it does not touch the system Hilbert space and thus, it can be applied 
to arbitrary (even time-dependent) impurity Hamiltonians and also to tunnel Hamiltonians $H_I$ with a global 
time-dependence. Furthermore, if the impurity is coupled to multiple reservoirs, the mapping can be applied to 
each reservoir {\it separately} such that it can be easily applied to the study of nonequilibrium scenarios (see 
Sec.~\ref{sec electronic MD beyond weak coupling}). 

In principle, the hope is that after the mapping, the problem is easier to tackle by using any preferred theoretical 
method. We here follow the idea to use a Markovian quantum ME for the extended system with redefined impurity 
Hamiltonian $\tilde H_\text{imp}$ (as shaded in grey in Fig.~\ref{fig RC mapping})~\cite{IlesSmithLambertNazirPRA2014, 
IlesSmithEtAlJCP2016, StrasbergEtAlNJP2016, NewmanMintertNazirPRE2017, StrasbergEspositoPRE2017}. To get a feeling for 
this approach, let us consider an initial Lorentzian SD of the form 
\begin{equation}\label{eq Lorentzian SD}
 J(\omega) = \frac{\Gamma\Delta^2}{(\omega-\omega_{0})^2 + \Delta^2},
\end{equation}
where $\Gamma$ describes the overall coupling strength and $\Delta$ the width of the Lorentzian centered around the 
resonance frequency $\omega_{0}$. We will indeed use this SD for our applications below. By applying 
Eqs.~(\ref{eq lambda_0}),~(\ref{eq E_1}) and~(\ref{eq mapping SD}), we obtain that 
the system couples to the RC with coupling strength $\lambda_0 = \sqrt{\Gamma\Delta/2}$ and energy $E_1 = \omega_0$, 
which is in turn coupled to a residual bath described by a SD of the form $J_1(\omega) = 2\Delta$. Thus, we see that 
the residual SD is completely flat, which is commonly believed to describe Markovian behaviour. In order to 
justify a ME approach, the redefined impurity should be additionally weakly coupled to the residual bath, which is 
exactly the case if the initial width $\Delta$ of the Lorentzian is small. Therefore, our method should be especially 
suited for the study of very structured SDs (e.g., described by sharp peaks) opposite to the wide-band limit. 
The equivalence for the particular example of a single quantum dot coupled to a bath with Lorentzian SD and a double 
quantum dot with flat SD was already noticed before~\cite{ElattariGurvitzPRA2000}, but we emphasize that our 
mapping is in general valid for any impurity system and SD. 

Even if the residual bath is not strictly Markovian or weakly coupled to the redefined system, one might 
hope that the ME including the RC provides nevertheless a convenient way to improve the accuracy of the results 
(compared to a conventional ME approach) as it takes a larger part of the model exactly into account. This is also 
supported by our benchmark in Appendix~\ref{app benchmark} where the coupling strength to the residual bath is rather 
moderate than weak. Moreover, if one is more interested in weak coupling than Markovianity, it is also possible to split 
the support of the SD into multiple intervals and to apply the RC mapping to each interval separately. This gives rise to 
more RCs, but also to a weaker coupling to the residual baths. Furthermore, it should be emphasized that the validity of 
any ME approach is likely to break down at very low temperatures. This, however, does not indicate a failure of the RC 
method itself. 

A benefit of the present approach is that it straightforwardly allows for a consistent thermodynamic 
interpretation~\cite{StrasbergEtAlNJP2016, NewmanMintertNazirPRE2017, StrasbergEspositoPRE2017, RestrepoEtAlArXiv2017}. 
In fact, due to the ME approach, we have direct access to the internal energy and entropy of the redefined impurity as 
well as to the energy and matter currents $I_E$ and $I_M$ from the residual reservoir, which are defined in 
Eqs.~(\ref{eq def IE ME}) and~(\ref{eq def IM ME}). If the system is coupled to multiple reservoirs $\nu$ at inverse 
temperature $\beta_\nu$ and with chemical potential $\mu_\nu$, it is straightforward to establish the validity of the 
nonequilibrium first law (energy balance) and second law (positivity of entropy production rate $\dot\Sigma$) of 
thermodynamics,
\begin{align}
 d_t\tilde E_\text{imp}(t)	&=	\sum_\nu \dot Q_\nu(t) + \dot W_\text{mech}(t) + \dot W_\text{chem}(t),	\\
 \dot\Sigma(t)			&\equiv	d_t\tilde S_\text{imp}(t) - \sum_\nu \beta_\nu\dot Q_\nu(t) \ge 0.	\label{eq 2nd law}
\end{align}
Here, $\tilde E_\text{imp}(t) = \mbox{tr}\{\tilde H_\text{imp}\tilde\rho_\text{imp}(t)\}$ is the internal energy and 
$\tilde S_\text{imp}(t) = -\mbox{tr}\{\tilde\rho_\text{imp}(t)\ln\tilde\rho_\text{imp}(t)\}$ is the entropy of the 
redefined impurity and the heat flows are given by $\dot Q_\nu(t) = I_E^{(\nu)} - \mu_\nu I_M^{(\nu)}$. In case of 
an explicit time-dependence, $\dot W_\text{mech}(t) = \mbox{tr}\{[d_t\tilde H_\text{imp}(t)]\tilde\rho_\text{imp}(t)\}$ 
denotes the mechanical work done on the system. Note that a clean derivation of the positivity of the entropy production, 
Eq.~(\ref{eq 2nd law}), requires a ME in Lindblad form although we have numerically observed no violation of positivity 
for our ME derived in Appendix~\ref{app ME}. 

The strategy to reformulate the laws of thermodynamics for an extended system incorporating a part of the previous 
bath has been suggested in Refs.~\cite{StrasbergEtAlNJP2016, NewmanMintertNazirPRE2017, StrasbergEspositoPRE2017, 
RestrepoEtAlArXiv2017}. It has the advantage that it avoids the difficulties faced with other methods such as 
Green's functions~\cite{EspositoOchoaGalperinPRL2015, EspositoOchoaGalperinPRB2015, BruchEtAlPRB2016, WhitneyArXiv2016, 
HaughianEspositoSchmidtPRB2018} or hierarchy of equations of motions~\cite{KatoTanimuraJCP2016, 
CerrilloBuserBrandesPRB2016}, where the interaction energy cannot be unambiguously assigned to the system or the bath when 
the system is not at steady state. 

We remark that the RC method allows to treat a larger class of initial conditions for the redefined impurity. To compare 
it with the conventional ME method~\cite{BreuerPetruccioneBook2002, SchallerBook2014, DeVegaAlonsoRMP2017} one has to 
choose the initial state of the RC to be equilibrated and decorrelated from the impurity state. In comparison with 
Green's function techniques~\cite{EspositoOchoaGalperinPRL2015, EspositoOchoaGalperinPRB2015, BruchEtAlPRB2016, 
WhitneyArXiv2016, HaughianEspositoSchmidtPRB2018} one usually assumes initially a global equilibrium state instead. 
Finally, we remark that by applying counting field techniques to the residual baths, nonequilibrium fluctuation relations 
such as those derived in Ref.~\cite{EspositoHarbolaMukamelRMP2009} also hold for our approach. 

To close this section, let us compare our method with the TEDOPA algorithm~\cite{PriorEtAlPRL2010, ChinEtAlJMP2010, 
WoodsEtAlJMP2014, WoodsCramerPlenioPRL2015, WoodsPlenioJMP2016, RosenbachEtAlNJP2016}, which can be straightforwardly 
extended to fermionic reservoirs~\cite{ChinEtAlJMP2010, WoodsEtAlJMP2014}. The goal of this method is to provide an exact 
mapping of the whole reservoir onto a semi-infinite chain of coupled fermions by using the theory of orthogonal 
polynomials. Then, one usually solves the whole system exactly by using DMRG methods~\cite{SchollwoeckRMP2005}. In 
principle, our method also allows by iterative application to obtain a semi-infinite chain of coupled fermions, but we 
believe that the strength of our method lies in the possibility to apply it step by step and to treat the transformed 
system by an approximate, yet simple and intuitive ME approach.

\section{The autonomous electronic Maxwell demon: A review}
\label{sec electronic MD review}

We here collect some recent results about autonomous Maxwell demons (MDs), especially within the context of electronic 
transport. We start by giving a simplified argument in Sec.~\ref{sec general argument} to underline why it is important 
to extend the study of autonomous MDs beyond the weak coupling regime. Sec.~\ref{sec electronic MD context} gives an 
overview over recent activities in the field with special attention paid to the electronic context. In 
Sec.~\ref{sec model} we then consider a particular important model, which we will numerically study using the theory of 
fermionic RCs in Sec.~\ref{sec electronic MD beyond weak coupling}. For comparison, Sec.~\ref{sec thermo ideal} finally 
reviews the thermodynamics of this model in the weak coupling regime and shows under which conditions the device can be 
viewed as an autonomous MD. The last two subsections also fix most of the notation and parameters 
eventually used in Sec.~\ref{sec electronic MD beyond weak coupling} to explore the non-Markovian regime.

\subsection{A general argument}
\label{sec general argument}

For readers unfamiliar with the working mechanism of an autonomous Maxwell demon, we here give a simplified description 
of it, which is in direct analogy with the device studied later on in this section. 

Let us start by specifying what we mean by an {\it autonomous MD}. We consider bipartite systems 
where one part can be understood as the controlled system and the other part assumes the role of the detector and 
controller by the physical interaction with the controlled system. The complete device should be autonomous in the sense 
that there is no time-dependence in the global Hamiltonian. Also external interventions by means of measurement and 
feedback control are forbidden, very similar to the idea to use ``coherent quantum control'' to simulate measurement based 
quantum control~\cite{WisemanMilburnPRA1994b, LloydPRA2000}. Thus, all physically relevant parts of the device are 
explicitly modeled. The device should also be a useful thermodynamic machine allowing, for instance, the extraction of 
work. These desiderata are also fulfilled by many other engines, but beyond that we especially require that:

(i)~The way information is processed in the device should be 
particularly transparent and these ``informational degrees of freedom'' are called the demon part of the device (whereas 
the rest is simply called the system); and 

(ii)~The energetics associated to the demon part should be negligible (though 
not necessarily strictly zero) compared to the energetics of the system itself, i.e., the device should be 
{\it information dominated}. 

For the moment, let us denote by $\Delta E_S$ ($\Delta E_D$) and $\tau_S$ ($\tau_D$) some characteristic energy scale 
and some typical relaxation time scale of the system (demon). Furthermore, we assume that the system (demon) has access 
to some heat reservoir at temperature $T_S$ ($T_D$). Whenever convenient, we will also use the notation of rates 
$\gamma_{S,D} = \tau_{S,D}^{-1}$ and inverse temperature $\beta_{S,D} = (k_B T_{S,D})^{-1}$ in our description. 
We emphasize that the relaxation rates $\gamma_{S,D}$ are within the conventional weak coupling approach proportional 
to the coupling strength between the system/demon and their respective reservoir; 
compare also with Appendix~\ref{app ME}. 

We start our analysis with the observation that an important feature of any MD is to extract work from fluctuations. 
Thus, in order to have non-negligible fluctuations in the system, we demand that 
\begin{equation}\label{eq cond fluctuations S}
 \Delta E_S \approx k_B T_S.
\end{equation}
Let us assume that the device delivers useful work and let us approximate the output power via 
\begin{equation}
 \dot W_\text{out} \approx \gamma_S \Delta E_S.
\end{equation}

In the {\it non}-autonomous version of MD, the second law for feedback control predicts that the maximum amount of work 
in the case of an ideal classical feedback controler is bounded by~\cite{ParrondoHorowitzSagawaNatPhys2015, 
StrasbergEtAlPRX2017} 
\begin{equation}\label{eq 2nd law feedback}
 \beta_S W_\text{out} = \beta_S \Delta E_S \lesssim \C I_{S:D},
\end{equation}
where $\C I_{S:D}$ denotes the mutual information between the system and the demon, 
\begin{equation}\label{eq def mutual info}
 \C I_{S:D} \equiv S(\rho_S) + S(\rho_D) - S(\rho_{SD}).
\end{equation}
Here, $\rho_{SD}$ denotes the density operator of the system and demon, $\rho_{S,D} = \mbox{tr}_{D,S}\{\rho_{SD}\}$ 
the marginal state and $S(\rho) \equiv -\mbox{tr}\{\rho\ln\rho\}$ the von Neumann entropy. Roughly speaking, $\C I_{S:D}$ 
quantifies the amount of correlations shared between the system and the demon and it plays an essential role in the field 
of information thermodynamics~\cite{ParrondoHorowitzSagawaNatPhys2015}. 
Based on this insight, the goal of an autonomous MD will also be to establish a strong correlation between the system and 
the demon part in order to harness it via some intrinsic feedback loop; thus, the demon part has to act like a 
detector. This implies that the demon must be able to react quickly enough to changes in the system, which naturally 
leads to the requirement 
\begin{equation}\label{eq requirement Ia}
 \gamma_D \gg \gamma_S ~~~ \text{(requirement Ia)},
\end{equation}
i.e., the demons typical time-scale $\tau_D$ is much smaller than $\tau_S$. This requirement inevitably tells us that 
in order to enhance the power output of the device by increasing $\gamma_S$, this necessarily implies a 
corresponding increase in $\gamma_D$, which in turn implies that the demon must be more strongly coupled to its own bath 
at temperature $T_D$. 

However, requirement~Ia is not enough to guarantee that the demon adapts with high probability to the correct state: 
given a certain state of the system, there should be a unique and stable state of the demon. Expressed differently, 
whereas we required the system to fluctuate relatively strongly, the demon should not fluctuate more than necessary. 
Therefore, for any system state the requirement of a reliable or precise demon translates into 
\begin{equation}\label{eq requirement Ib}
 \frac{\Delta E_D}{k_B T_D} \gg 1 ~~~ \text{(requirement Ib)}.
\end{equation}

Requirement~Ia and~Ib are linked to our desideratum (i) mentioned initially. Point~(ii) is fulfilled by the requirement 
\begin{equation}\label{eq requirement II}
 \Delta E_S \gg \Delta E_D ~~~ \text{(requirement II)}.
\end{equation}
From requirement Ib and II and condition~(\ref{eq cond fluctuations S}) it also follows immediately that 
\begin{equation}
 k_B T_S \approx \Delta E_S \gg \Delta E_D \gg k_B T_D.
\end{equation}
Hence, the heat reservoir of the demon must necessarily be much colder than the heat reservoir of the system. 

To conclude, the above argumentation shows us that an autonomous MD needs to be carefully tuned. The fact that the demon 
acts like a detector requires it to be fast and precise and thus, to be much more strongly coupled to its own bath than 
the system. On top of that, a small energy consumption requires the demon to have access to a low entropy (or low 
temperature) reservoir. Both requirements, strong coupling and low temperature, challenge the usual range of validity 
of commonly employed perturbative approaches. The rough estimates given here should be also compared with the analysis 
in Sec.~\ref{sec thermo ideal}.

\subsection{Maxwell's demon in the electronic context}
\label{sec electronic MD context}

The central idea of an electronic MD is to find some feedback mechanism, which shovels particles 
against a chemical gradient instead of a thermal gradient as in the traditional thought-experiment of Maxwell. 
The big advantage of this setup comes from recent technological advances, which allow to measure and manipulate 
single electrons in quantum dots, see, e.g.,~\cite{FujisawaEtAlScience2006, FlindtEtAlPNAS2009, KoskiEtAlPNAS2014, 
HartmannEtAlPRL2015, ThierschmannEtAlNatNanotech2015, KoskiEtAlPRL2015, WagnerEtAlNatNano2017, ChidaEtAlNatComm2017}. 
Direct measurement and manipulation of individual phonons, which are predominantly responsible for thermal transport, 
is much harder instead. Thus, there have been many theoretical studies how to use charge fluctuation in mesoscopic 
conductors to extract work via some feedback loop, either in an autonomous way~\cite{Datta2008, StrasbergEtAlPRL2013, 
MosshammerBrandesPRB2014, KutvonenEtAlSciRep2015, LebedevEtAlPRA2016, RoselloLopezPlateroPRB2017, 
MertBozkurtPekertenAdagideliArXiv2017} or not~\cite{SchallerEtAlPRB2011, AverinMottonenPekolaPRB2011, 
KishGranqvistPloS12012, EspositoSchallerEPL2012, BergliGalperinKopninPRE2013, StrasbergEtAlPRE2014, StrasbergEtAlPRX2017, 
EngelhardtSchallerNJP2018, SchallerEtAlArXiv2017}. 

In the following, we will focus on the autonomous MD introduced in Ref.~\cite{StrasbergEtAlPRL2013}. 
Globally, it resembles a thermoelectric device based on two Coulomb coupled quantum dots~\cite{SanchezBuettikerPRB2011, 
SanchezJordanNanotechnology2015}. However, by a careful fine-tuning of 
the parameters it is possible to show that it resembles the phenomenological electronic MD introduced 
in Ref.~\cite{SchallerEtAlPRB2011} and thus, it has a particularly transparent interpretation in terms of information 
flows~\cite{EspositoSchallerEPL2012, StrasbergEtAlPRL2013, HorowitzEspositoPRX2014, KutvonenEtAlSciRep2015}. Moreover, 
very similar experimental realizations were 
reported~\cite{HartmannEtAlPRL2015, ThierschmannEtAlNatNanotech2015, KoskiEtAlPRL2015, ChidaEtAlNatComm2017} 
though it remains unclear whether it is possible to reach the ideal information-dominated regime. 

\begin{figure}
 \centering\includegraphics[width=0.22\textwidth,clip=true]{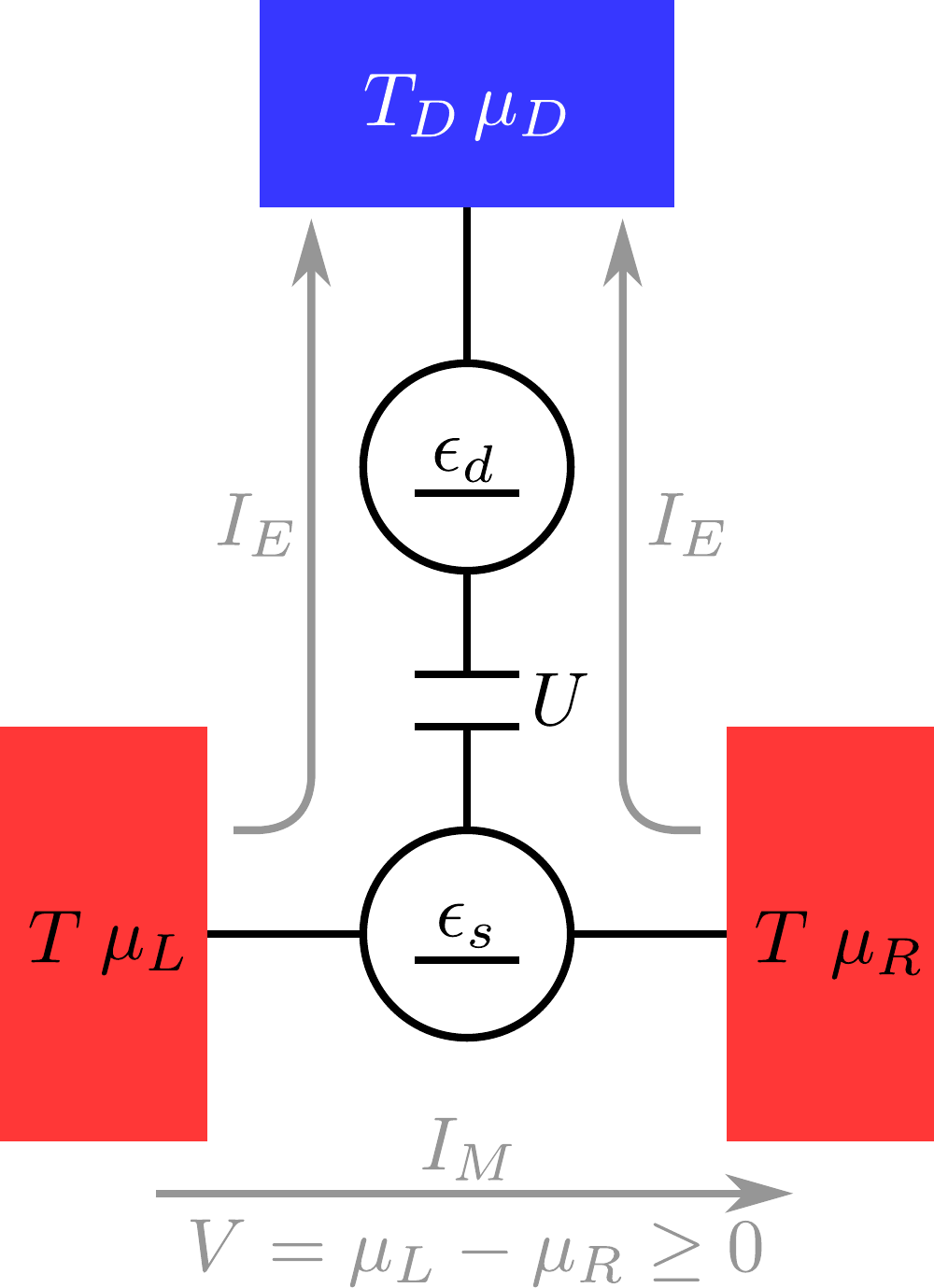}
 \label{fig setup} 
 \caption{Sketch of the autonomous MD device. Two quantum dots (black circles) with on-site energies 
 $\epsilon_s$ and $\epsilon_d$ interact capacitatively with strength $U$. The demon dot is tunnel-coupled to an 
 electronic reservoir kept at temperature $T_D$ and chemical potential $\mu_D$. The system dot is tunnel-coupled to 
 two reservoirs kept at the same temperature $T$, but with different chemical potentials $\mu_L - \mu_R = V \ge 0$. 
 Grey arrows indicate the two independent currents $I_E$ and $I_M$ pointing in the direction where they are defined 
 to be positive. The goal of the demon is to reverse the electric current (i.e., $I_M < 0$) with the smallest possible 
 imbalance in the energy flow through the system. }
\end{figure}

\subsection{Model}
\label{sec model}

The model is schematically depicted in Fig.~\ref{fig setup}. The Hamiltonian of the impurity (system and demon) reads 
\begin{equation}\label{eq sys Ham}
 H_\text{imp} = \epsilon_s d_s^\dagger d_s + \epsilon_d d_d^\dagger d_d + U d_s^\dagger d_s d_d^\dagger d_d
\end{equation}
where $d_{s/d}^\dagger$ ($d_{s/d}$) is a fermionic creation (annihilation) operator for the system/demon dot. 
The Hamiltonian describes two dots with on-site energies $\epsilon_s$ and $\epsilon_d$ and Coulomb interaction 
$U$, which we will treat exactly with our method. Note that there are no electrons tunneling between the dots. 

The non-interacting reservoirs $\nu\in\{L,R,D\}$ are modeled as free fermions as in Eq.~(\ref{eq H imp}) with Hamiltonian 
$H_R^{(\nu)} = \sum_k \epsilon_{k\nu}c^\dagger_{k\nu} c_{k\nu}$. 
The reservoirs $L$ and $R$ are tunnel-coupled to the system quantum dot via the Hamiltonian 
\begin{equation}
 H_I^{(\nu)} =  \sum_k\left(t_{k\nu} d_s c_{k\nu}^\dagger + t_{k\nu}^* c_{k\nu} d_s^\dagger\right) ~~~ (\nu\in\{L,R\}),
\end{equation}
whereas the reservoir $D$ is tunnel-coupled to the demon dot via 
\begin{equation}
 H_I^{(D)} = \sum_k\left(t_{kD} d_d c_{kD}^\dagger + t_{kD}^* c_{kD} d_d^\dagger\right).
\end{equation}
The total Hamiltonian then reads $H_\text{tot} = H_\text{imp} + \sum_{\nu\in\{L,R,D\}}(H_I^{(\nu)} + H_R^{(\nu)})$. 
Within the weak coupling approach each reservoir is assumed to be well described by an equilibrium distribution according 
to a temperature $T_\nu$ and chemical potential $\mu_\nu$. Below, we will set for simplicity $T_L = T_R \equiv T$ and 
$\mu_L = \epsilon_s + V/2$ and $\mu_R = \epsilon_s - V/2$ where $V = \mu_L - \mu_R \ge 0$ denotes the bias voltage 
across the system. 

The SD $J^{(\nu)}(\omega)$ of each reservoir $\nu$ is modeled by a Lorentzian as in Eq.~(\ref{eq Lorentzian SD}) with 
coupling strength $\Gamma_\nu$, peak width $\Delta_\nu$ and resonance frequency $\omega_{0\nu}$. In the following, we 
will set $\Gamma_L = \Gamma_R \equiv \Gamma_S$ and $\Delta_L = \Delta_R \equiv \Delta_S$, i.e., the system is coupled 
with equal strength to the left and to the right reservoir and only the position of the peaks will be different. 

\begin{figure*}
 \centering\includegraphics[width=0.95\textwidth,clip=true]{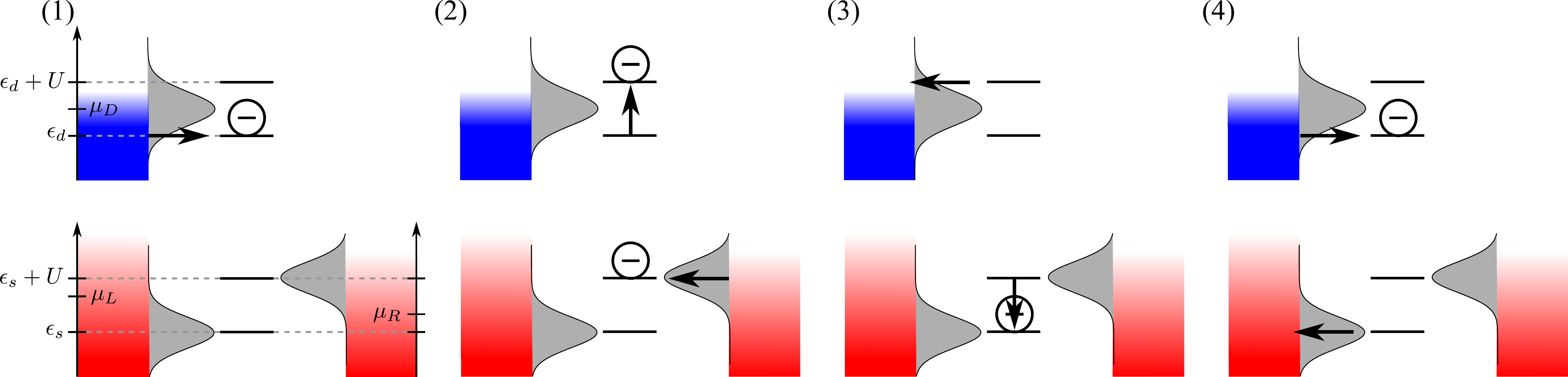}
 \label{fig trajectory} 
 \caption{A possible stochastic trajectory which becomes likely in the MD regime as explained at the end of 
 this caption. In contrast to 
 Fig.~\ref{fig setup}, we here sketched the transition frequencies $\epsilon_s,\epsilon_d,\epsilon_s+U$ and 
 $\epsilon_d+U$ in the dots. The Fermi distribution in the baths are shown as density profiles where the darkness of 
 the color is proportional to the occupancy of that energy. To complete the picture we also sketched 
 the Lorentzian SDs of each reservoir whose product with the Fermi function determines the transition rates 
 (see Appendix~\ref{app ME}). The trajectory starts with a filled dot of the demon but an empty dot of the system (1). 
 If an electron wants to enter the system, it needs an energy $\epsilon_s + U$ and, although the Fermi factor is smaller 
 for the right reservoir, this is overcompensated by the imbalance of the Lorentzian tunneling rates (2). After an 
 electron has entered the system, it pushes the electron in the demon dot to a higher energy above the chemical 
 potential such that it instantaneously jumps out of the dot due to requirement~(\ref{eq limit fast demon}) (3). 
 Finally, and again due to the imbalance of the Lorentizian tunneling rates, the electron in the system dot leaves to 
 the left bath and the demon dot gets refilled (4). Overall, one electron was transfered against the bias by harnessing 
 the correlation between the system and the demon dot and by transfering an energy amount $U$ from the hot to the cold 
 reservoir. }
\end{figure*}

\subsection{Thermodynamics in the idealized limit}
\label{sec thermo ideal}

In the idealized picture the device is weakly coupled to three Markovian thermal reservoirs as depicted in 
Fig.~\ref{fig setup}. Within this approach the dynamics of the system is governed by a master equation (ME) with rates 
obtained from Fermi's golden rule, see Ref.~\cite{StrasbergEtAlPRL2013} for the full rate ME. Alltogether the system 
works as a thermoelectric device~\cite{SanchezBuettikerPRB2011, StrasbergEtAlPRL2013}. If we set $T_D < T$, the device 
is able to convert an energy current $I_E \equiv -I_E^{(D)} > 0$ flowing into the reservoir $D$ into an electric current 
$I_M \equiv I_M^{(L)} <0$ flowing against the bias $V$ provided that the SDs $J^{(\nu)}(\omega)$ and other parameters are 
choosen appropriately [for a definition of energy and matter currents see Eqs.~(\ref{eq def IE ME}) 
and~(\ref{eq def IM ME})]. Because the device is out of equilibrium, it produces entropy at a rate 
\begin{equation}\label{EQ:entprod_total}
 \dot\Sigma = \beta V I_M + (\beta_D - \beta)I_E \ge 0.
\end{equation}
It is also possible to give compact analytical expressions for $I_M$ and $I_E$ in terms of the steady 
state~\cite{SanchezBuettikerPRB2011, StrasbergEtAlPRL2013}, 
but for our purposes it suffices to note the following proportionality relation: 
\begin{equation}\label{eq proportionality IS}
 I_M \sim \Gamma_S.
\end{equation}
This means that the power output of the device is directly proportional to the coupling strength $\Gamma_S$ in the 
weak coupling regime and hence, it is desirable to choose $\Gamma_S$ as large as possible. Furthermore, we add that one 
can associate a second law to the local dynamics of the system only, which reads~\cite{HorowitzEspositoPRX2014}
\begin{equation}\label{eq 2nd law local}
 \dot\Sigma_S = \beta VI_M - \beta I_E - \dot{\C I}_S \ge 0.
\end{equation}
Here, $\dot{\C I}_S$ is the {\it flow} of mutual information between the system and the demon (for a precise definition 
see Ref.~\cite{HorowitzEspositoPRX2014}). Eq.~(\ref{eq 2nd law local}) shows that the ability to shovel electrons against 
the bias, $\beta V I_M < 0$, can be influenced by energetic ($\beta I_E$) as well as entropic ($\dot{\C I}_S$) 
contributions. 

More importantly for our discussion, there is a clean limit in which the system can be viewed as an autonomous MD
and in which the {\it reduced} dynamics of the system quantum dot coincides with the ideal, non-autonomous MD studied 
in Ref.~\cite{SchallerEtAlPRB2011}. In this regime the stochastic trajectory shown in Fig.~\ref{fig trajectory} becomes 
likely and transport against the bias possible. We will shortly review here this limit for completeness, more details can 
be found elsewhere~\cite{StrasbergEtAlPRL2013, HorowitzEspositoPRX2014}. 
The three different limits we need to take are the following (compare also with Sec.~\ref{sec general argument}): 
\begin{enumerate}
 \item[1a] {\it Fast demon:} First, the demon needs to be relatively fast in order to adapt quickly enough to the system 
 state such that it is guaranteed that the correct feedback loop is applied. This is ensured by demanding that 
 \begin{equation}\label{eq limit fast demon}
  \Gamma_D \gg \Gamma_S,
 \end{equation}
 implying that the demon is much more strongly coupled to its reservoir than the system, cf.~Eq.~(\ref{eq requirement Ia}). 
 Within the formal limit $\Gamma_D/\Gamma_S\rightarrow\infty$, the demon dot can be adiabatically eliminated and a closed 
 effective description for the system dot alone emerges. In the strict limit the flow of mutual information becomes 
 exactly identical to $\dot{\C I}_S = -\beta_U I_E$ so that the effective second law coincides with the true second law, 
 $\dot\Sigma_S = \dot\Sigma$. 
 \item[1b] {\it Precise demon:} In order to use the demon as a detector, it should also be precise in the sense that it is 
 as correlated as possible with the system since this increases the mutual information [compare with 
 Eq.~(\ref{eq 2nd law feedback})]. Optimal results can be achieved by choosing the chemical potential of the demon 
 such that $\epsilon_d = \mu_D - U/2$ and by requiring [cf.~Eq.~(\ref{eq requirement Ib})]
 \begin{equation}\label{eq limit precise demon}
  \beta_D U\gg 1.
 \end{equation}
 Then, in the strict limit of $\Gamma_D/\Gamma_S\rightarrow\infty$ and $\beta_D U\rightarrow\infty$ 
 the demon dot is occupied if and only if the system dot is empty and vice versa implying a perfect (anti-) correlation 
 between the system and the demon. It should be noted, however, that this strict limit also implies a diverging entropy 
 production rate~(\ref{EQ:entprod_total}). Additional details concerning the use of such setups as detection devices 
 can be found in Refs.~\cite{SchallerKiesslichBrandesPRB2010, SchulenborgEtAlPRB2014}.
 \item[2] {\it Maxwell demon:} Within the fast and precise demon limit, the demon can be reliably interpreted as a 
 detector, yet it still disturbs the energetic balances of the system at the order of $U$. Not very surpisingly, 
 the MD limit consists of demanding that 
 \begin{equation}\label{eq limit Maxwell demon}
  \beta V \approx 1, ~~~ \beta(V+U) \approx 1, ~~~ \frac{U}{\epsilon_s} \ll 1,
 \end{equation}
 which is in agreement with Eqs.~(\ref{eq cond fluctuations S}) and~(\ref{eq requirement II}) if we note that the energy 
 current through the system is roughly proportional to $\epsilon_S I_M$ in the limit of negligible $U$. Thus, in this 
 limit the difference in energy transfered from the right to the left reservoir (which is given by the flow of energy in 
 the detector bath $I_E$) becomes immeasurably small compared to the energetic current $\epsilon_S I_M$. In this 
 regime the term $\beta I_E$ becomes negligible and the second law reads 
 \begin{equation}
  \dot\Sigma = \dot\Sigma_S = \beta VI_M - \dot{\C I}_S \ge 0,
 \end{equation}
 i.e., the ability to shovel electrons against the bias is purely entropy (or information) dominated. 
\end{enumerate}

The above three limits are, however, not sufficient to shovel electrons against the bias. In order to achieve this, we also 
need to break the left-right symmetry of the device even in the absence of any bias ($V=0$). This is most easily done by 
choosing different SDs $J^{(L)}(\omega)$ and $J^{(R)}(\omega)$ fulfilling $J^{(L)}(\epsilon_s) > J^{(L)}(\epsilon_s + U)$ 
and $J^{(R)}(\epsilon_s) < J^{(R)}(\epsilon_s + U)$ as depicted in Fig.~\ref{fig trajectory}. More specifically, we choose 
for our Lorentzian SDs 
\begin{equation}
 \omega_{0L} = \epsilon_s, ~~~ \omega_{0R} = \epsilon_s + U.
\end{equation}
The imbalance in the SDs is then quantified by 
\begin{equation}\label{eq imbalance SD}
 \frac{J^{(L)}(\epsilon_s)}{J^{(L)}(\epsilon_s+U)} = \frac{J^{(R)}(\epsilon_s+U)}{J^{(R)}(\epsilon_s)} = 1+\frac{U^2}{\Delta_S^2},
\end{equation}
which was previously~\cite{StrasbergEtAlPRL2013} described by the parameter $e^\delta$. 

Furthermore, we choose to fix the following parameters for all upcoming numerical calculations: We set the on-site 
energies equal $\epsilon_s=\epsilon_d=\epsilon$, the inverse temperature to $\beta\epsilon = 1$, and we choose a small, 
but finite positive bias $V = 0.01\epsilon$. In addition, we set $U = 0.015\epsilon$ such that it is small compared to 
the system energy, see Eq.~(\ref{eq limit Maxwell demon}). The SD of the demon is peaked around 
$\omega_{0D} = \mu_D = \epsilon_d + U/2$ and we choose $\Delta_D = 0.01\epsilon$. 
The coupling strength of the demon is set to $\Gamma_D = 100\Gamma_S$ ensuring that the demon is fast enough. 

Thus, there are three free parameters left: (1) the variance $\Delta_S$ of the Lorentzian SD of the left and right 
reservoir controls the imbalance of the SDs [see Eq.~(\ref{eq imbalance SD})] thereby having influence on the power output 
and the validity of the Markov approximation; (2) the tunneling rate $\Gamma_S$ 
which controls the work output [see Eq.~(\ref{eq proportionality IS})] and (via $\Gamma_D = 100\Gamma_S$) the strength of 
the coupling between the demon dot and its reservoir; and (3) the temperature ratio $\beta_D/\beta$ influencing the 
precision of the demon [see Eq.~(\ref{eq limit precise demon})] and thus, the correlation with the system.

\begin{figure*}
 \centering\includegraphics[width=0.75\textwidth,clip=true]{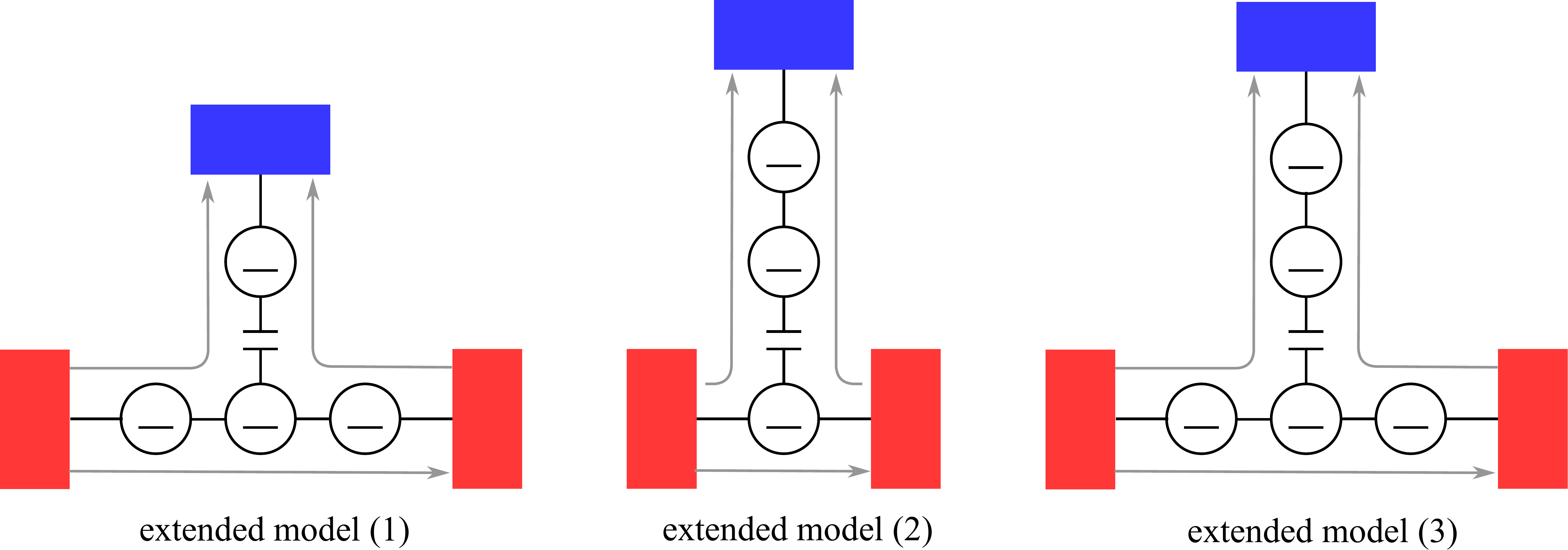}
 \label{fig extended models} 
 \caption{Sketch of the geometry of the extended models studied in the text. Each circle represents a single fermionic 
 site (``quantum dot''). The tunnel coupling is indicated by straight lines and the Coulomb interaction by a capacitor. 
 The three reservoirs as well as the energy and electric current are also indicated as in Fig.~\ref{fig setup}. }
\end{figure*}

\section{Electronic Maxwell demon beyond weak coupling}
\label{sec electronic MD beyond weak coupling}

In this section we report our main findings which are based on numerical comparison of the ideal (i.e., weakly coupled, 
Markovian and high temperature) model, see Sec.~\ref{sec thermo ideal}, with the extended RC approach able to go beyond 
these limitations.\footnote{Numerical results were obtained using a Mathematica notebook, which is available upon 
request from the corresponding author. To compute marginal states, as needed for Eq.~(\ref{eq def mutual info}), we used 
the algorithm ``Partial trace of a multiqubit system'' by Mark S.~Tame. } 
We will focus on the steady state behaviour only and compare three important quantities. 

First, the electric current $I_M$ flowing through the system dot from the left to the right reservoir. 
It is directly proportional to (minus) the work output of the device and is 
therefore an important quantifier for the overall thermodynamic performance. 

Second, we look at the relative energetic imbalance $|I_E/I_E^{(L)}|$ where $I_E = -I_E^{(D)}$ denotes the energy flow 
into the demon reservoir and $I_E^{(L)}$ the energy flow from reservoir $L$ [for a definition of energy and particle 
currents, see Eqs.~(\ref{eq def IE ME}) and~(\ref{eq def IM ME})]. If this quantity is large, the device works in an 
energy dominated regime, whereas it goes to zero in the Maxwell demon limit~(\ref{eq limit Maxwell demon}). 

Third, we will evaluate the mutual information $\C I_{S:D}$ between the system and the demon dot. It reveals key insights 
about the question whether the working mechanism of the device can be seen as some implicit measurement and 
feedback loop, which needs large correlations. Note that for our setup $0 \le \C I_{S:D} \le 2\ln 2$, but the maximum 
value is attained only for a pure and maximally entangled state. The classical limit is instead 
$\C I_{S:D}^\text{cl} \le \ln 2$ such that values relatively close to this limit signify already strong correlations 
in our noisy setup. In our numerical studies we have found that it suffices to restrict the plots of the mutual 
information to the interval $[0,\ln 2]$, although this does not a priori imply that there are no quantum correlations 
present.

\subsection{Extended models}

The validity of the standard ME approach as sketched in Appendix~\ref{app ME} is limited by the coupling strength to the 
reservoirs, the degree of non-Markovianity and the temperature of the reservoirs. In principle, all three limitations 
are challenged by the electronic MD model from Sec.~\ref{sec electronic MD review}. We therefore compare it with 
three different extended models: 

(1) The working mechanism of our device requires a breaking of the left-right symmetry, which was achieved by 
choosing different SDs $J^{(L)}(\omega)$ and $J^{(R)}(\omega)$ peaked around $\epsilon_s$ and $\epsilon_s+U$ respectively. 
In the ideal case, where the peaks are very narrow, the SDs act like perfect electron filters increasing the 
thermoelectric performance. Quite problematically, however, a strongly peaked SD is usually associated with strong 
non-Markovianity~\cite{ZedlerEtAlPRB2009, DeVegaAlonsoRMP2017}. In order to capture the non-Markovian behaviour of the 
left and right reservoir, we introduce fermionic RCs $C_l^{(\dagger)}$ and $C_r^{(\dagger)}$ for the left and the right 
reservoir in what we call ``model 1''. For the spectral densities centered at $\omega_{0,L}=\epsilon_s$ and 
$\omega_{0,R}=\epsilon_s+U$, the resulting redefined impurity Hamiltonian reads 
\begin{equation}
 \begin{split}\label{eq Ham mod 1}
  \tilde H_\text{imp}^{(1)}	=&~	H_\text{imp} + \epsilon_s C_l^\dagger C_l + (\epsilon_s+U)C_r^\dagger C_r	\\
				&+	\sqrt{\frac{\Gamma_S \Delta_S}{2}}\left(C_l^\dagger d_s + d_s^\dagger C_l + C_r^\dagger d_s + d_s^\dagger C_r\right)
 \end{split}\,,
\end{equation}
and the coupling to the residual left and right reservoir is described by a flat SD $J_1^{(\nu)}(\omega) = 2\Delta_S$, 
$\nu\in\{L,R\}$. A sketch of the setup is shown in Fig.~\ref{fig extended models}. 
We remark that the inclusion of additonal quantum dots to model peaked SDs acting as energy filtes has been 
used elsewhere, too~\cite{SanchezBuettikerPRB2011, SanchezPlateroBrandesPRB2008, SanchezThierschmannMolenkampNJP2017}. 

(2) Putting the problem of non-Markovianity aside, the most pressing problem is that the demon dot must be relatively 
strongly coupled to a low temperature heat reservoir. This problem becomes more pronounced if we want to enhance the 
power output by increasing $\Gamma_S$. Model (2) therefore introduces a fermionic RC $C_d^{(\dagger)}$ for the demon bath 
in order to study those effects. When we use $\omega_{0,D}=\epsilon_d+U/2$, the resulting redefined impurity Hamiltonian 
reads 
\begin{equation}\label{eq Ham mod 2}
 \begin{split}
  \tilde H_\text{imp}^{(2)}	=&~	H_\text{imp} + \left(\epsilon_d + \frac{U}{2}\right) C_d^\dagger C_d	\\
				&+	\sqrt{\frac{\Gamma_D\Delta_D}{2}}(C_d^\dagger d_d + d_d^\dagger C_d)
 \end{split}
\end{equation}
and the coupling to the residual demon reservoir is described by a flat SD $J_1^{(D)}(\omega) = 2\Delta_D$. 
A sketch of the setup is again shown in Fig.~\ref{fig extended models}. 

(3) In order to capture both problems, model 3 finally uses a fermionic RC for all reservoirs, see 
Fig.~\ref{fig extended models}. Within our framework this will provide the ultimate test for the simplied model from 
Sec.~\ref{sec thermo ideal}. The redefined Hamiltonian is given by the sum of Eq.~(\ref{eq Ham mod 1}) 
and~(\ref{eq Ham mod 2}) without counting $H_\text{imp}$ twice, 
\begin{equation}
 \tilde H_\text{imp}^{(3)} = \tilde H_\text{imp}^{(1)} + \tilde H_\text{imp}^{(2)} - H_\text{imp}.
\end{equation}
The SDs of the residual reservoirs $L, R$ and $D$ 
are given by $2\Delta_S, 2\Delta_S$ and $2\Delta_D$, respectively.

\begin{figure}
 \centering\includegraphics[width=0.41\textwidth,clip=true]{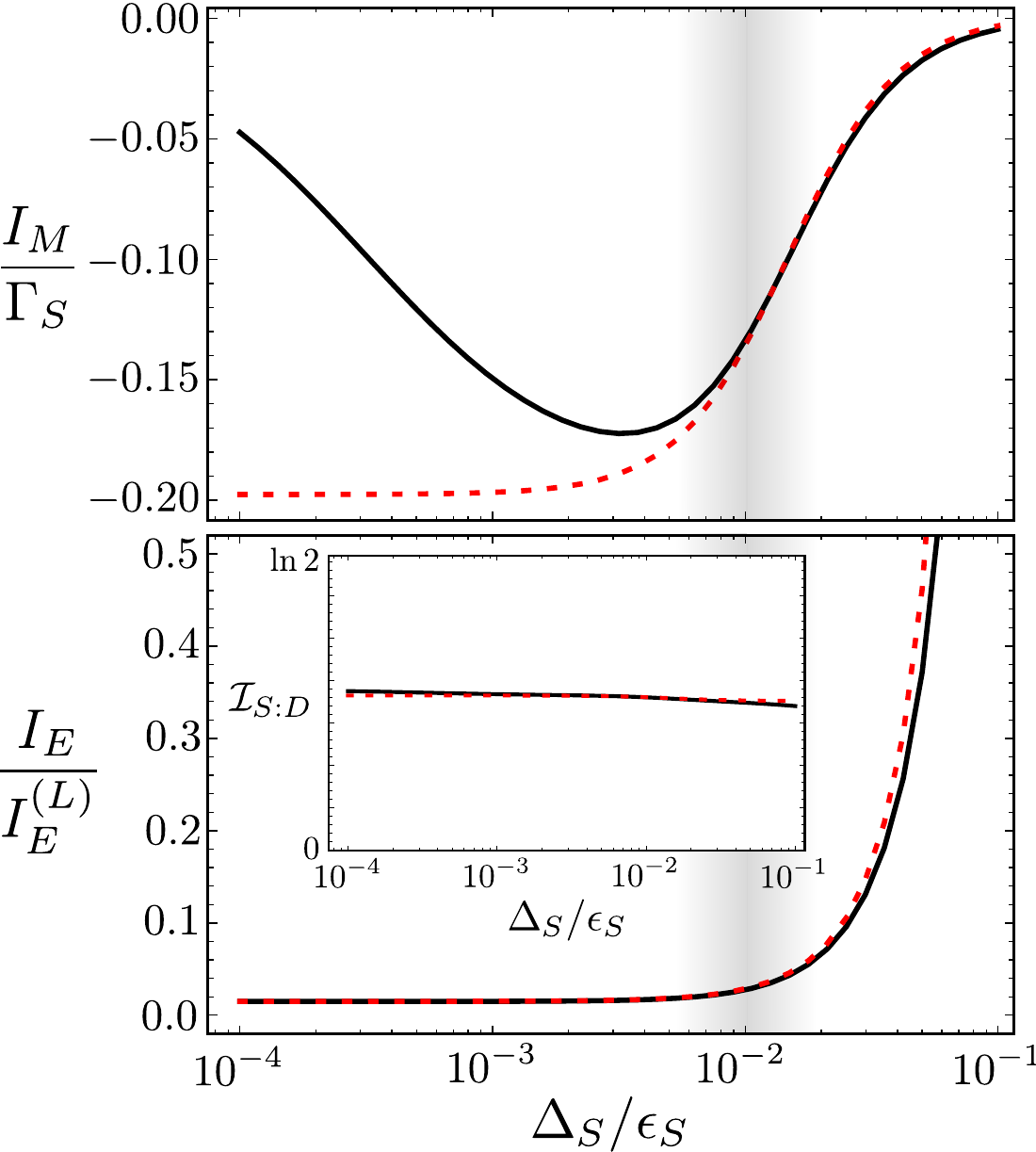}
 \label{fig plot 1} 
 \caption{Plot of the dimensionless quantities $I_M/\Gamma_S$ (top), $I_E/I_E^{(L)}$ (bottom) and $\C I_{S:D}$ (inset) 
 versus the dimensionless sharpness of the peak of the left and right SDs $\Delta_S/\epsilon_s$ (note the logarithmic 
 scale). Results using the extended model (1) are shown with a solid line, the dotted red line refers to the DQDMD 
 description from Sec.~\ref{sec thermo ideal}. The shaded grey area indicates the 
 region where the DQDMD treatment and the MD interpretation is valid. To remain in 
 the strict weak coupling regime,  we chose $\Gamma_S = 10^{-5}\epsilon_s$. Furthermore, $\beta_D/\beta = 300$. }
\end{figure}

\subsection{Numerical results}

In what follows, we will compare the results of our extended models with the original MD treatment
exposed in Sec.~\ref{sec thermo ideal}. We will refer to the latter as double quantum dot Maxwell demon (DQDMD) treatment 
and use red, dashed lines in the plots. The regime where the simple double quantum dot treatment {\it and} the MD 
interpretation are valid are shaded in grey in the plots.

\subsubsection{RCs for the system reservoirs}

We start with the extended model (1) to answer the question how narrowly peaked can we choose the SDs $J^{(\nu)}(\omega)$ 
before the Markovian description from Sec.~\ref{sec thermo ideal} breaks down? The answer is shown in 
Fig.~\ref{fig plot 1}. It clearly shows that there is an optimum value for the sharpness of the peak, which can be found 
by maximizing the power output $-VI_M$ while still retaining a valid Markovian description. 

For $\Delta_S\rightarrow0$ the RC description indeed predicts $I_M\rightarrow0$ because the coupling to the residual 
baths is directly proportional to $\Delta_S$. It is worth to emphasize that the naive ME treatment of 
Sec.~\ref{sec thermo ideal} predicts a monotonically {\it increasing} power output in the limit $\Delta_S\rightarrow0$ 
in strong contrast to the actual behaviour. We conclude that one has to be very careful if one chooses strongly preaked 
SDs (sometimes called ``spectral filters'') to increase the power output of a thermodynamic device. 

In the opposite limit, $\Delta_S\rightarrow\infty$, the engine also 
breaks down because the SDs become essentially flat and the left-right symmetry remains unbroken. Note that the amount 
of mutual information between the system and the demon is almost unaffected by the shape of the SD as expected. 

Although the simple DQDMD description becomes inaccurate below a certain critical threshold 
$\Delta_S^*/\epsilon_s \approx 0.01$, the plot also shows (shaded grey region) that a Markovian description is valid even 
if the SD is not completely flat but slightly structured [at the critical value, the imbalance~(\ref{eq imbalance SD}) is 
for the choosen numerical parameters roughly $1.18$]. Finally, we note that the regime, where the Markovian description 
is valid while at the same time the energetic imbalance is very small (less than $5\%$), is restricted to a very narrow 
window around $\Delta^*_S/\epsilon_s$ demonstrating the careful finetuning needed to ensure a working autonomous MD. 

\begin{figure}
 \centering\includegraphics[width=0.41\textwidth,clip=true]{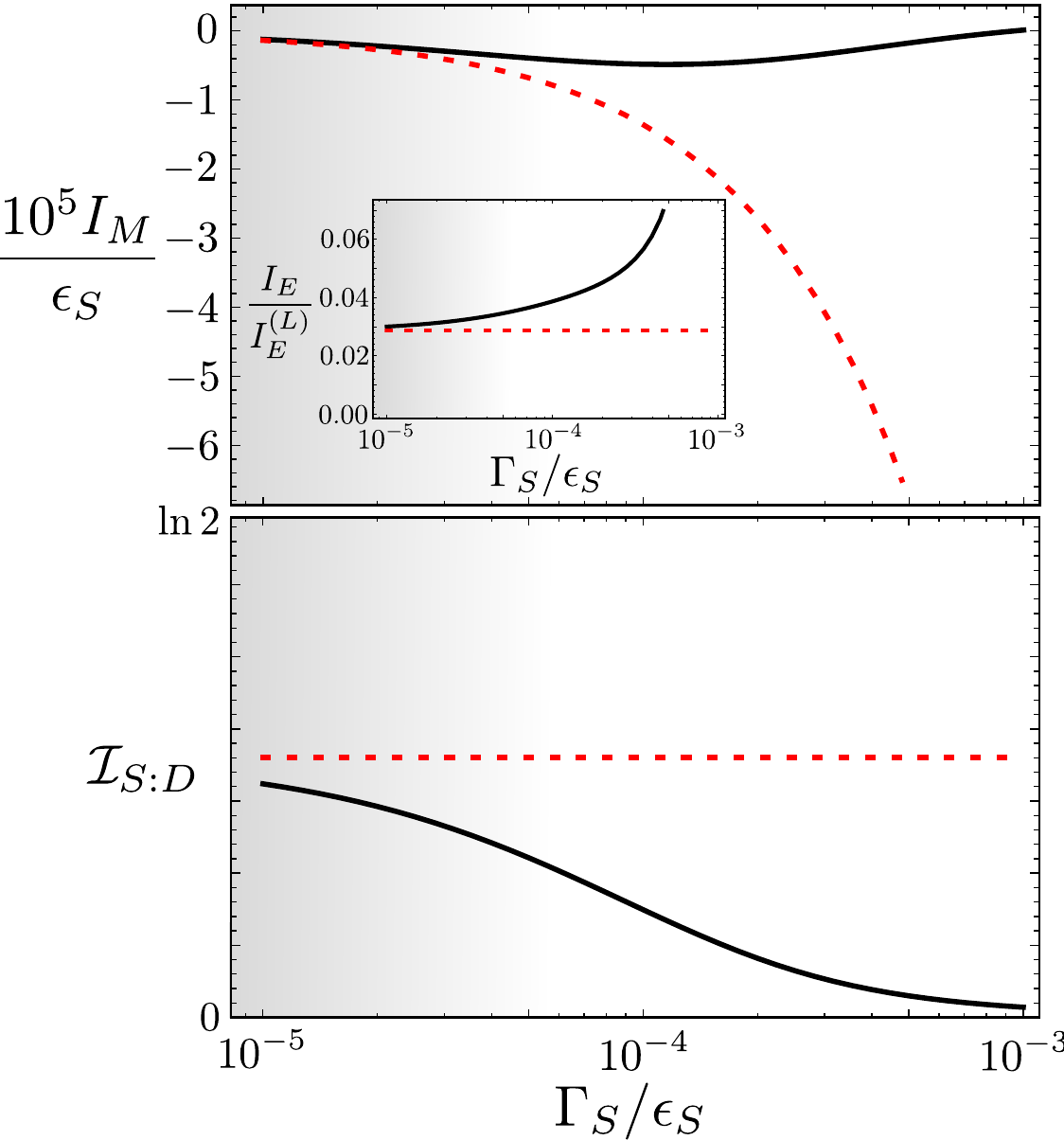}
 \label{fig plot 21} 
 \caption{Plot of our three relevant quantities versus the dimensionless coupling strength $\Gamma_S$ of the system 
 (note the logarithmic scale again) for the extended model (2) (solid line) and the DQDMD treatment (dotted red line). 
 The shaded grey area indicates the region where the DQDMD treatment and the MD interpretation is valid. 
 Parameters as in Fig.~\ref{fig plot 1} with $\Delta_S/\epsilon_s = 0.01$. }
\end{figure}

\begin{figure}
 \centering\includegraphics[width=0.41\textwidth,clip=true]{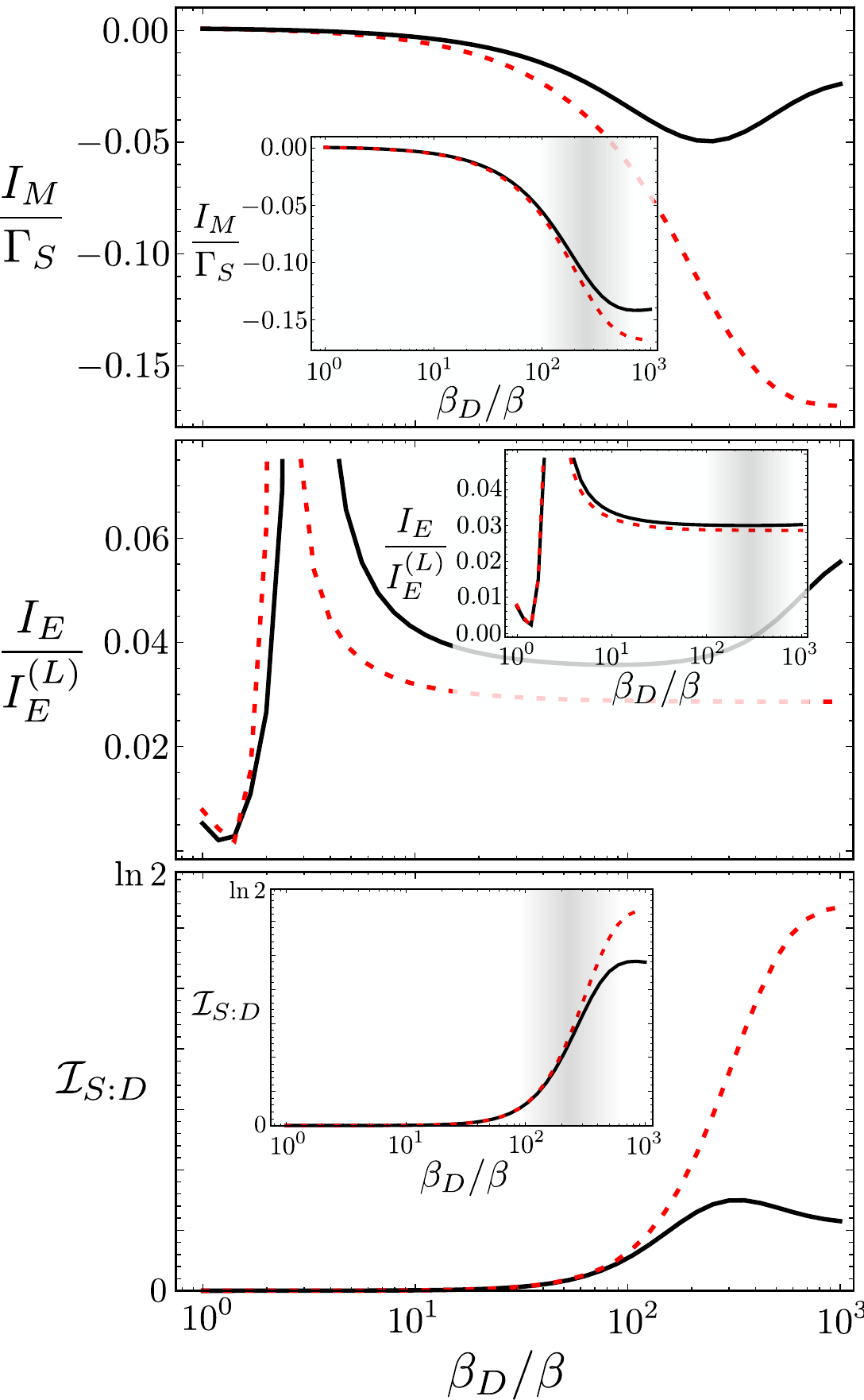}
 \label{fig plot 22} 
 \caption{Plot of our three relevant quantities versus the dimensionless inverse temperature $\beta_D/\beta$ 
 of the demon reservoir (in logarithmic scale) for $\Gamma_S=10^{-4} \epsilon_s$ (main plots) and 
 $\Gamma_S=10^{-5} \epsilon_s$ (insets). The rest is as in Fig.~\ref{fig plot 21}. }
\end{figure}

\subsubsection{RC for the demon reservoir}

Numerical results associated to model (2) concerning the question what happens to the demon in the strong coupling and 
low temperature regime are shown in Figs.~\ref{fig plot 21} and~\ref{fig plot 22}. 

\begin{figure*}
 \centering\includegraphics[width=0.99\textwidth,clip=true]{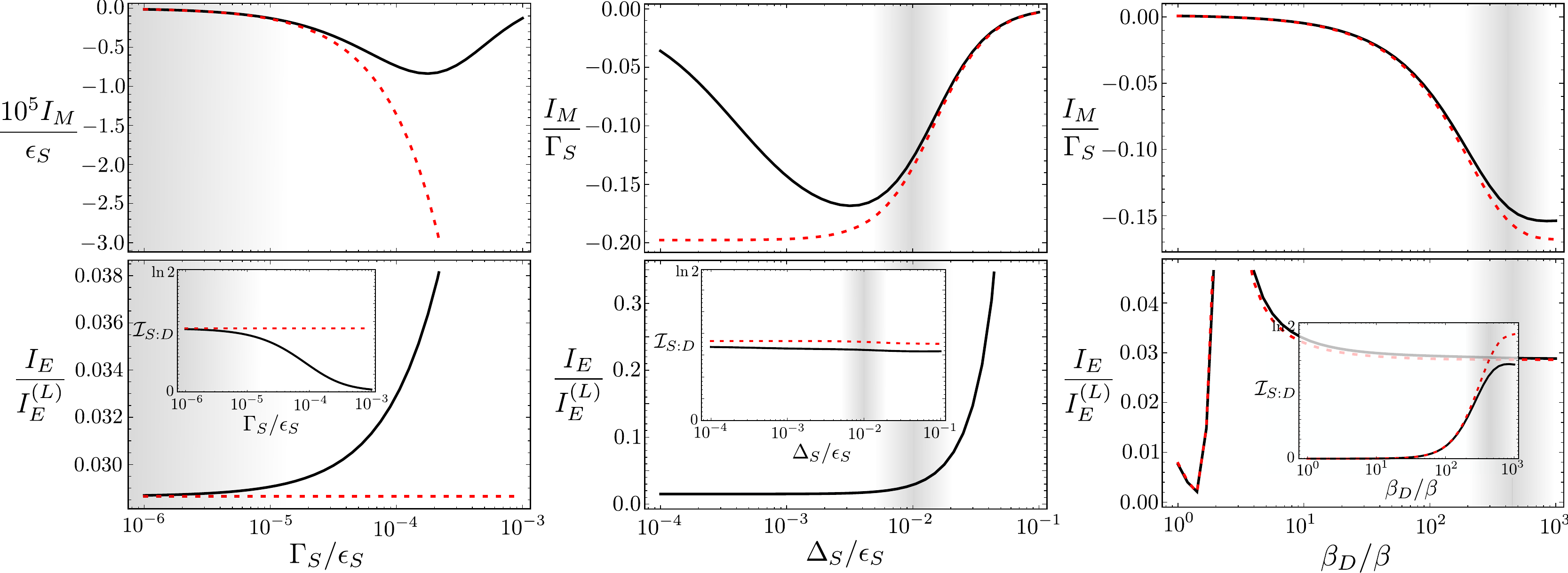}
 \label{fig plot 3} 
 \caption{Repetition of the plots shown before for the extended model (3). In all plots we chose 
 $\Gamma_S = 10^{-5}\epsilon_s$, $\Delta_S = 0.01\epsilon_s$ and $\beta_D/\beta = 300$ unless that parameter is varied in 
 the respective plot. The grey shaded area indicates the region, in which the ideal MD interpretation holds. }
\end{figure*}

First, Fig.~\ref{fig plot 21} shows our three relevant quantities as a function of $\Gamma_S$, which (in the weak 
coupling regime) is directly proportional to the power output of the device 
[Eq.~(\ref{eq proportionality IS})] and influences the demon coupling strength via our choice $\Gamma_D = 100 \Gamma_S$. 
Numerical parameters are the same as in Fig.~\ref{fig plot 1} with the choice $\Delta_S = \Delta_S^* = 0.01\epsilon_s$. 
As expected, the plot of the electric current $I_M$ demonstrates that the DQDMD treatment is only valid for very 
small coupling strength $\Gamma_S< 10^{-4}\epsilon_s$ (shaded grey region), beyond that the demon fails to work. 
Also the plot of the relative energy 
imbalance $I_E/I_E^{(L)}$ shows that we are leaving the MD regime of negligible energy consumption for larger 
$\Gamma_S$ because more transport channels are opening up. More interestingly, however, is the plot of the mutual 
information $\C I_{S:D}$, which reveals the physical 
reason why the demon fails. As explained in Sec.~\ref{sec electronic MD review}, the working mechanism is based on a 
strong correlation between the demon and the system due to the Coulomb interaction and a careful tuning of the demon's 
reservoir. But for stronger coupling the electron in the demon dot gets more and more correlated with its reservoir than 
with the system, i.e., it becomes more and more delocalized. The only way to counter balance this behaviour is by 
increasing the Coulomb interaction $U$, but this increases the energy imbalance pushing us away from the MD regime. 
This is the ultimate reason why the MD is limited to the weak coupling situation and thus, due to the required time-scale 
separation $\Gamma_D \gg \Gamma_S$, to very low power output. 

To complement the analysis, Fig.~\ref{fig plot 22} shows the same quantities for varying inverse temperature $\beta_D$ 
for relatively strong coupling strength $\Gamma_S=10^{-4}\epsilon_s$ and, shown as insets, for 
$\Gamma_S=10^{-5}\epsilon_s$, which was used in Fig.~\ref{fig plot 1}. As expected, for larger $\Gamma_S$ we see stronger 
deviations from the ideal values 
confirming our previous observation. In addition, Fig.~\ref{fig plot 22} demonstrates two more important features. 
First, the device works better for lower temperatures because the formation of correlations is hindered if the demon is 
subjected to more thermal noise. Second, this figure also shows that a DQDMD description is usually only valid 
at high temperatures, whereas more sophisticated methods are needed for lower temperatures. We stress once more that, 
though the RC method allows to treat lower temperatures, its validity also does not extend down to zero temperature 
$T_D\rightarrow0$. 

\subsubsection{RCs for all reservoirs}

Finally, numerical results using model~(3) are shown in Fig.~\ref{fig plot 3}. It is our most accurate analysis and 
combines model~(1) and~(2) and the plots demonstrate that the results, which we have drawn above in a separate analysis, 
hold true also in the complete picture. Furthermore, although the simple DQDMD picture from 
Sec.~\ref{sec thermo ideal} fails for most parameters as expected, it also agrees well if we pay careful attention to its 
range of validity. Thus, the analysis of the ideal MD~\cite{StrasbergEtAlPRL2013}, despite the various 
limits involved, remains qualitatively and quantitatively true for a narrow parameter window.

\section{Summary}
\label{sec conclusions}

In this article we have developed the theory of fermionic RCs, which provides a tool to extend the range 
of validity of the usual ME, especially for very structured, i.e., strongly non-Markovian, SDs. The benefit of our 
approach is that the ME approach still allows to treat interactions in the system exactly, it can be straightforwardly 
applied to nonequilibrium situations and has a transparent thermodynamic interpretation. One drawback of the method 
is that not every initial SD can be mapped to an effectively weakly coupled and Markovian situation such that the 
application of the ME has to be justified on a case-by-case study. 
Another drawback comes from the use of a ME itself, 
which becomes invalid at very low temperatures. Nevertheless, we believe that the fermionic RC mapping has the potential 
to find widespread application in quantum transport as a simple and transparent tool to treat structured SDs. 

As a particular application we then considered a thermoelectric device, which can be interpreted as an autonomous MD for 
a specific range of parameters. Previous analyses in the field of information thermodynamics were  done 
in the idealized weak coupling and Markovian regime, an exception being Ref.~\cite{WalldorfJauhoKaasbjergPRB2017}. 
As we have argued in Sec.~\ref{sec general argument} and as Ref.~\cite{StrasbergEtAlPRL2013} explicitly 
shows, MD lives in a parameter regime where the use of these idealized assumptions becomes increasingly questionable. 
We here addressed for the first time systematically the question of what happens to the performance of an autonomous 
MD if we relax the weak coupling and Markovian assumption
(also see Ref.~\cite{SchallerEtAlArXiv2017} for the study of a {\it non}-autonomous MD in the strong coupling regime). 

Our numerical results clearly convey two messages: First, we have proven that it is indeed possible to reach the 
idealized MD regime, even if one uses a more sophisticated method, which is able to take into account strong coupling 
and non-Markovian effects. Thus, MD is not a mere hypothetical being, but can be found in actual physical systems. 
On the other hand, our article also indicates that the possible parameter regime of an ideal MD is very narrow and 
necessarily limited to low power output. It therefore still remains a challenge to find out to what extend MD will play 
a role in actual, practically useful devices, where already the use of a model Hamiltonian of the form~(\ref{eq sys Ham}) 
amounts to a strong assumption as it neglects, e.g., spin and vibrational degrees of freedom as well 
as multiple charges on a single quantum dot. 

{\it Acknowledgements. } 
Important initial discussions with Tobias Brandes on the fate of Maxwell's demon in the strong-coupling and low 
temperature regime have led to these investigations. GS acknowledges financial support by the DFG (SCHA 1646/3-1, 
SFB 910, and GRK 1558), TLS acknowledges support from the National Research Fund Luxembourg (ATTRACT 7556175), 
and PS and ME by the European Research Council project NanoThermo (ERC-2015-CoG Agreement No. 681456). 
GS and PS furthermore acknowledge support of the WE-Heraeus foundation for supporting the 640 WE-Heraeus seminar. 


\bibliography{/home/philipp/Documents/references/books,/home/philipp/Documents/references/open_systems,/home/philipp/Documents/references/thermo,/home/philipp/Documents/references/info_thermo,/home/philipp/Documents/references/general_QM}

\appendix
\section{Derivation of the master equation}
\label{app ME}

In this appendix we briefly state the essential steps to derive the quantum ME used in the main text. We start by 
focusing on an impurity (also often called the ``system'') coupled to a single reservoir with Hamiltonian 
$H_\text{tot} = H_\text{imp} + H_I + H_R$. Because within the weak coupling approach there is no direct influence 
between the different reservoirs, we can simply add the contributions of them at the end. More detailed derivations of 
MEs can be found elsewhere~\cite{BreuerPetruccioneBook2002, SchallerBook2014, DeVegaAlonsoRMP2017, 
EspositoHarbolaMukamelRMP2009}. Note that $H_\text{imp}$, $H_I$ and $H_R$ 
are arbitrary and left unspecified here. For the MEs used in Sec.~\ref{sec electronic MD beyond weak coupling} one would 
indeed need to replace $H_\text{imp}$ by $\tilde H_\text{imp}$. 

Our starting point is the second order Liouville-von Neumann equation in the interaction 
picture after performing the Markovian approximation 
\begin{equation}
 \begin{split}
  d_t\tilde\rho(t) = -\int_0^\infty d\tau	&	\mbox{tr}_B\left\{\tilde H_I(t)\tilde H_I(t-\tau)\tilde\rho(t)R_0\right.	\\
						&	\left.- \tilde H_I(t)\tilde\rho(t)R_0\tilde H_I(t-\tau) + h.c.\right\}
 \end{split}
\end{equation}
where $\tilde A(t) \equiv e^{i(H_\text{imp}+H_R)t} A e^{-i(H_\text{imp}+H_R)t}$ denotes operators in the interaction 
picture and $R_0 = e^{-\beta(H_R-\mu N_R)}/Z$ describes the equilibrium density operator of the reservoir 
with $N_R$ being the particle number operator of the reservoir. 

For our purposes it suffices to consider an interaction Hamiltonian of the form 
\begin{equation}
 H_I = d\sum_k t_kc_k^\dagger + \sum_k t_k^* c_k d^\dagger
\end{equation}
where $d$ is an arbitrary fermionic system operator. 
We denote the eigensystem and transition frequencies of $H_\text{imp}$ as 
\begin{equation}
 H_\text{imp} = \sum_k E_k|k\rl k|, ~~~ \omega_{kl} \equiv E_k - E_l
\end{equation}
such that we can write the system coupling operator in the interaction picture as 
\begin{equation}
 \tilde d(t) = \sum_{k,l} e^{i\omega_{kl}t} d_{kl} |k\rl l|, ~~~ d_{kl} \equiv \langle k|d|l\rangle. 
\end{equation}
Then, after introducing the SD $J(\omega) = 2\pi\sum_k |t_k|^2\delta(\omega-\epsilon_k)$ of the bath and after 
moving out of the interaction picture, we obtain terms like 
\begin{widetext}
\begin{align}
 & e^{-iH_\text{imp}t}\mbox{tr}_B\{\tilde H_I(t)\tilde H_I(t-\tau)\tilde\rho(t)R_0\}e^{iH_\text{imp}t}	\nonumber	\\
 & ~~~ = \int_{-\infty}^\infty d\omega \frac{J(\omega)}{2\pi}\sum_{k,l} \left\{e^{i(\omega+\omega_{kl})\tau} f(\omega) d d^*_{kl} |l\rl k|\rho(t) + e^{-i(\omega+\omega_{kl})\tau} [1-f(\omega)] d^\dagger d_{kl} |k\rl l|\rho(t)\right\},	\\
 & e^{-iH_\text{imp}t}\mbox{tr}_B\{\tilde H_I(t)\tilde\rho(t)R_0\tilde H_I(t-\tau)\}e^{iH_\text{imp}t}	\nonumber	\\
 & ~~~ = \int_{-\infty}^\infty d\omega \frac{J(\omega)}{2\pi}\sum_{k,l} \left\{e^{i(\omega+\omega_{kl})\tau} [1-f(\omega)] d\rho(t) d^*_{kl} |l\rl k| + e^{-i(\omega+\omega_{kl})\tau} f(\omega) d^\dagger \rho(t) d_{kl} |k\rl l|\right\}
\end{align}
\end{widetext}
and their Hermitian conjugate. Here, 
$f(\epsilon_k) = \mbox{tr}_B\{c_k^\dagger c_k R_0\} = [e^{\beta(\epsilon_k-\mu)}+1]^{-1}$ denotes the Fermi distribution. 

To evaluate the integral over $\tau$, we then use 
\begin{equation}\label{eq principal value}
 \int_0^\infty d\tau e^{\pm i\omega\tau} = \pi\delta(\omega) \pm i\C P\frac{1}{\omega}
\end{equation}
and {\it neglect} the imaginary principal value part in the following (cf.~Appendix~\ref{app benchmark}). 
After this step the full ME can be written as 
\begin{equation}
 \begin{split}\label{eq ME numerics}
  d_t\rho(t)	=&	-i[H_\text{imp},\rho(t)] + [\chi^\dagger\rho(t),d] + [d^\dagger,\rho(t)\chi]	\\
		&+	[\theta\rho(t),d^\dagger] + [d,\rho(t)\theta^\dagger].
 \end{split}
\end{equation}
where we introduced the operators 
\begin{align}
 \chi	&\equiv	\sum_{k,l}\frac{J(\omega_{lk})}{2}f(\omega_{lk}) d_{kl}|k\rl l|,	\\
 \theta	&\equiv	\sum_{k,l}\frac{J(\omega_{lk})}{2}[1-f(\omega_{lk})] d_{kl}|k\rl l|.
\end{align}
The ME~(\ref{eq ME numerics}) is ready for numerical implementation. Note that we did not perform the commonly employed 
secular approximation, which guarantees a Lindblad form of the generator~\cite{BreuerPetruccioneBook2002, 
SchallerBook2014, DeVegaAlonsoRMP2017, EspositoHarbolaMukamelRMP2009}, 
but often predicts unphysical results especially for more complex systems, see also Appendix~\ref{app benchmark}. 

If we include coupling to multiple baths characterized by a distinct chemical potential or temperature, 
we can simply add up the contribution of each bath separately to the ME. The result is then 
\begin{equation}\label{eq ME numerics multiple reservoirs}
 d_t\rho(t) = -i[H_\text{imp},\rho(t)] + \sum_\nu \C L_\nu\rho(t)
\end{equation}
with the dissipator 
\begin{equation}
 \C L_\nu\rho \equiv [\chi_\nu^\dagger\rho,d_\nu] + [d_\nu^\dagger,\rho\chi_\nu] + [\theta_\nu\rho,d_\nu^\dagger] + [d_\nu,\rho\theta_\nu^\dagger].
\end{equation}
The energy and particle currents into bath $\nu$ are then given by 
\begin{align}
 I_E^{(\nu)}	&=	\mbox{tr}\{H_\text{imp}\C L_\nu\rho(t)\},	\label{eq def IE ME}	\\
 I_M^{(\nu)}	&=	\mbox{tr}\{N_\text{imp}\C L_\nu\rho(t)\},	\label{eq def IM ME}
\end{align}
where $N_\text{imp}$ is the particle number operator of the impurity.

\section{Example and benchmark: Single electron transistor}
\label{app benchmark}

\begin{figure*}
 \centering\includegraphics[width=0.70\textwidth,clip=true]{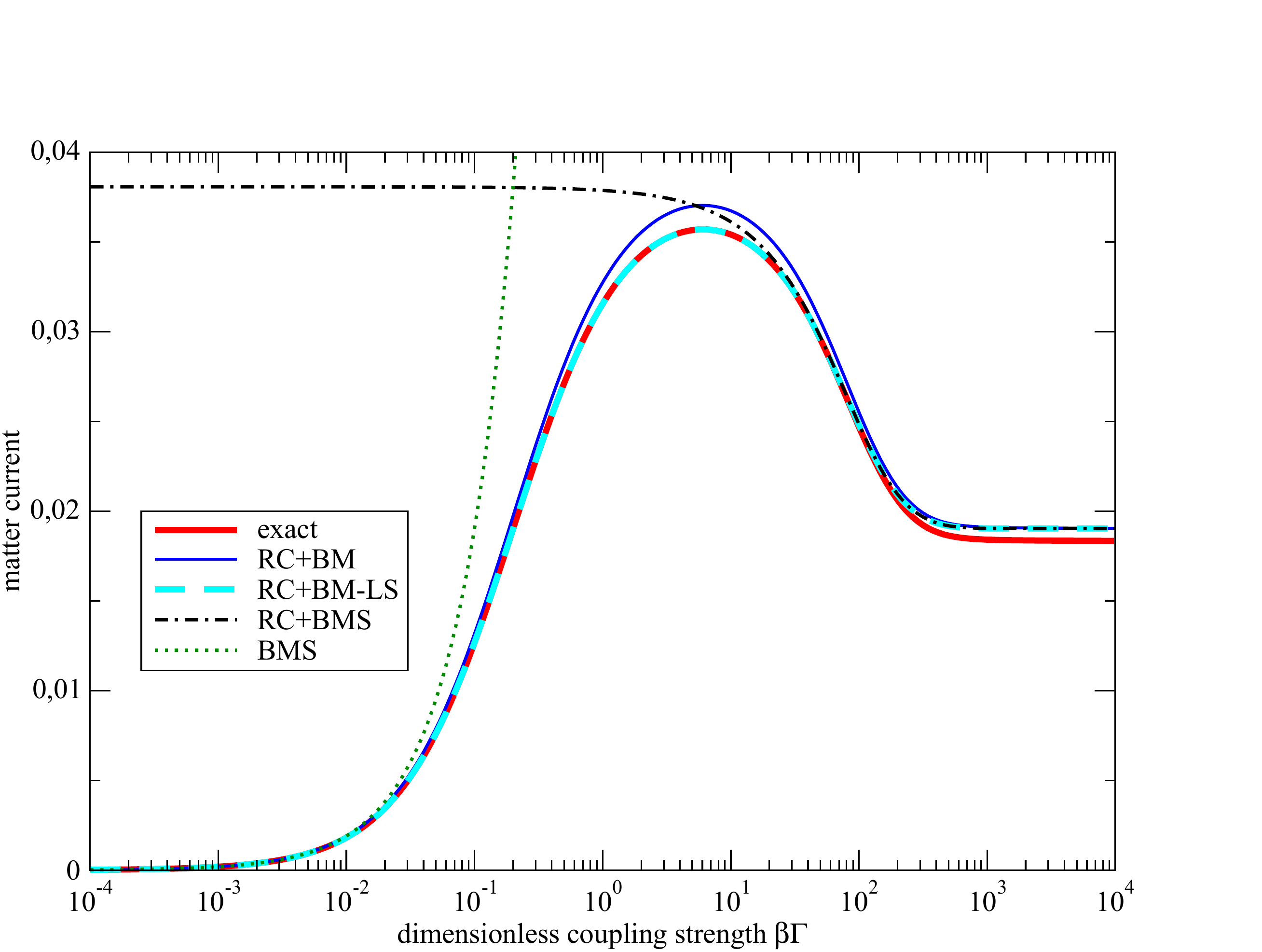}
 \label{fig comp SET 1} 
 \caption{Plot of the matter current over the dimensionless coupling strength $\beta\Gamma$ in logarithmic scale computed 
 using different methods: the naive rate equation approach (thin dotted green), the exact solution (thick red), 
 and the RC approach based on different levels of approximation for the ME:  The Born-Markov treatment of 
 Eq.~(\ref{eq ME numerics}) including the Lamb shift terms from Eq.~(\ref{eq principal value}) (RC+BM, thin solid dark 
 blue curve), Eq.~(\ref{eq ME numerics}) neglecting the Lamb shift (RC+BM-LS, dashed bold light blue curve), and 
 Eq.~(\ref{eq ME numerics}) with an additional secular approximation (BMS, thin dash-dotted solid black). Parameters are 
 $\Delta = 0.1 \epsilon, \omega_0 = \epsilon, \beta\epsilon = 1$ and $\mu_L = \epsilon = -\mu_R$. }
\end{figure*}

To benchmark our approach we consider the possibly simplest fermionic transport setup usually called a single electron 
transistor (SET): a spinless quantum dot with on-site energy $\epsilon$ coupled to two fermionic baths $\nu\in\{L,R\}$. 
In this case we have $H_\text{imp} = \epsilon d^\dagger d$ and we model the contact to the baths again by a Lorentzian of 
the form~(\ref{eq Lorentzian SD}), which we assume to be the same for the left and right reservoir. 

\begin{figure*}
 \centering\includegraphics[width=0.70\textwidth,clip=true]{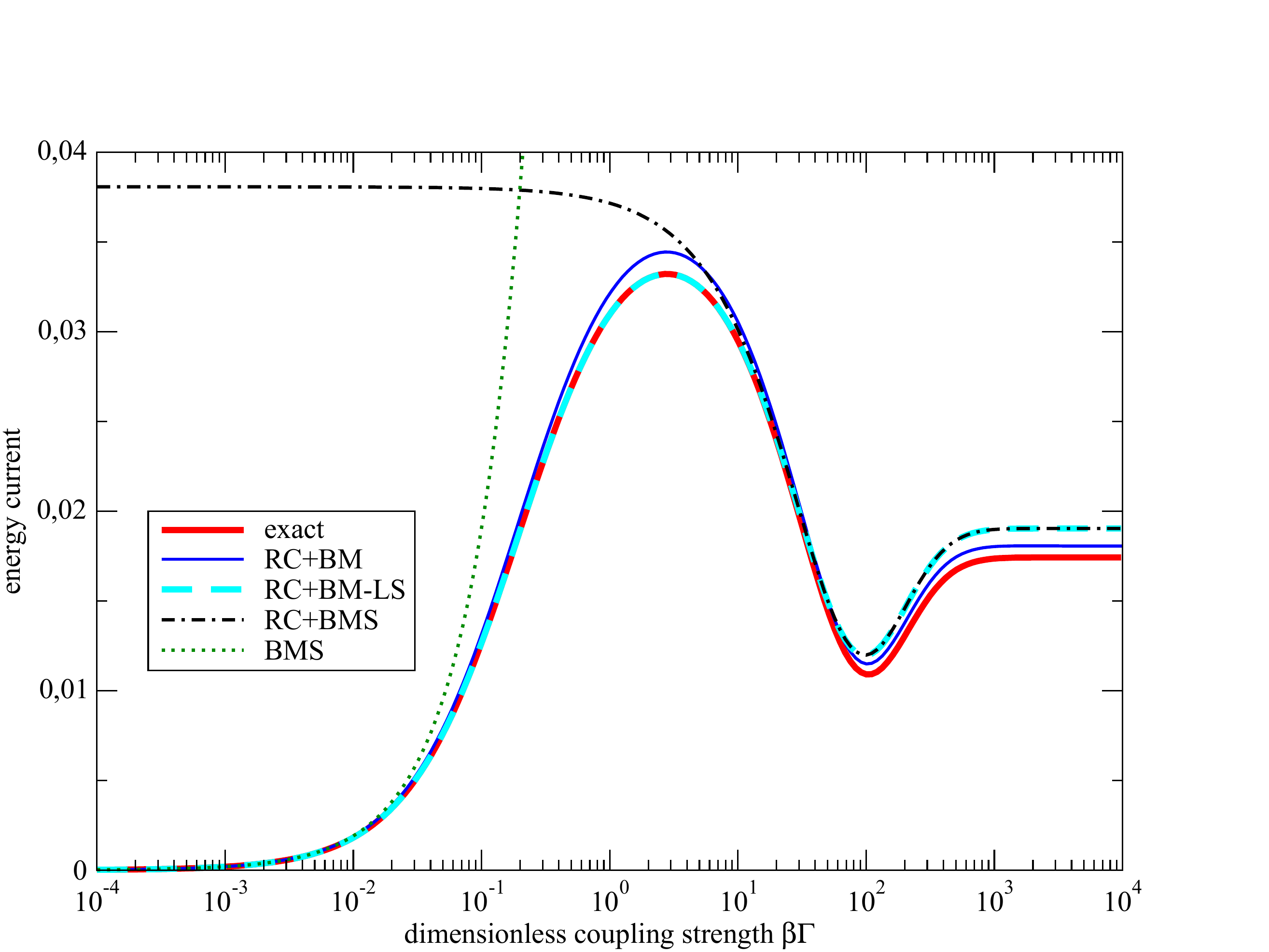}
 \label{fig comp SET 2} 
 \caption{Plot of the energy current instead of the matter current with the same parameters as in 
 Fig.~\ref{fig comp SET 2}. }
\end{figure*}

Figs.~\ref{fig comp SET 1} and~\ref{fig comp SET 2} now compare the energy and particle current through the SET using 
three different methods. 
The first (dotted green) is based on a standard rate equation for the probability to find the quantum dot empty 
or filled (i.e., the ME from Appendix~\ref{app ME} applied to $H_\text{imp} = \epsilon d^\dagger d$). 
The second method 
makes use of a single RC mapping individually applied to each reservoir and yields the Hamiltonian 
\begin{equation}
 \tilde H_\text{imp} = H_\text{imp} + \sum_\nu\left(\lambda_0dC_{1\nu}^\dagger + \lambda_0^*C_{1\nu}d^\dagger + E_1C_{1\nu}^\dagger C_{1\nu}\right),
\end{equation}
which now consists of three serially coupled quantum dots. The second method then treats the residual baths as weakly 
coupled and Markovian by using the ME treatment from Appendix~\ref{app ME}. In addition, it allows for different 
perturbative treatments by either including (solid dark-blue line) or neglecting (dashed light-blue line) Lamb shift 
terms or by performing an additional secular approximation on top (dash-dotted black line). Finally, the model also admits 
an exact solution for the matter and energy current which reads (solid red line; compare with, e.g., 
Ref.~\cite{ToppBrandesSchallerEPL2015}) 
\begin{align}
 I_M	&=	\int_{-\infty}^\infty \frac{2d\omega}{\pi} \frac{J_{L}(\omega)J_{R}(\omega)[f_L(\omega) - f_R(\omega)]}{[J_{L}(\omega)+J_{R}(\omega)]^2 + 4[\omega - \epsilon - \Sigma(\omega)]^2},	\nonumber	\\
 I_E	&=	\int_{-\infty}^\infty \frac{2d\omega}{\pi} \frac{\omega J_{L}(\omega)J_{R}(\omega)[f_L(\omega) - f_R(\omega)]}{[J_{L}(\omega)+J_{R}(\omega)]^2 + 4[\omega - \epsilon - \Sigma(\omega)]^2}.	\nonumber
\end{align}
Here, $\Sigma = \Sigma_L + \Sigma_R$ denotes the Lamb shift 
\begin{equation}
 \begin{split}
  \Sigma_\nu(\omega)	&\equiv	\C P \int \frac{d\omega'}{2\pi} \frac{J_{\nu}(\omega')}{\omega-\omega'}	\\
			&=	\frac{\Gamma\Delta(\omega-\omega_0)}{2[(\omega-\omega_0)^2+\Delta^2]} = \frac{\omega-\omega_0}{2\Delta}J_{\nu}(\omega).
 \end{split}
\end{equation}

As we can see, the RC method based on the ME~(\ref{eq ME numerics}) (i.e., {\it without} secular approximation) gives an 
excellent agreement with the exact result for a wide parameter regime. Generally, we see that over all coupling strengths, 
including or neglecting the Lamb shift has little effect. For intermediate coupling strengths, including the Lamb shift 
terms is even reducing the agreement with the exact solution, whereas for the ultrastrong coupling regime it is improving 
the agreement, as is particularly visible in the energy current. More importantly, however, we see that the often employed 
secular approximation fails completely in the weak to intermediate parameter regimes. These reasons justify the use of the 
ME we derived in Appendix~\ref{app ME}. 

It is nevertheless important to remark that the RC method is also limited. First of all, in order to justify using 
a weak coupling ME for $\tilde H_\text{imp}$, the width $\Delta$ of the Lorentzian must be small enough because it is 
directly proportional to the coupling strength of the RC with the residual bath. Second, even for small $\Delta$ 
our approach is still based on the use of a ME invalidating our results for very low temperature. For instance, the 
differential conductance of the SET $\lim_{V\rightarrow0} \frac{dI_M}{dV}$ computed with the RC method completely fails 
to reproduce the exact results for $T\rightarrow0$ (not shown here for brevity).

\end{document}